\documentclass[prd, showpacs, superscriptaddress, floatfix, nofootinbib]{revtex4}

\usepackage[dvips]{graphicx}
\usepackage{epsf}
\usepackage{amsmath}
\usepackage{amssymb}

\usepackage{graphicx}
\usepackage{dcolumn}
\usepackage{bm}
\def\be{\begin{equation}}
\def\ee{\end{equation}}
\def\bea{\begin{eqnarray}}
\def\eea{\end{eqnarray}}

\usepackage{color}

\newcommand{\mpl}{m_{_\mathrm{Pl}}}
\newcommand{\Mp}{M_{_\mathrm{Pl}}}

\newcommand{\gsimm}{\raise.3ex\hbox{$>$\kern-.75em\lower1ex\hbox{$\sim$}}}
\newcommand{\lsimm}{\raise.3ex\hbox{$<$\kern-.75em\lower1ex\hbox{$\sim$}}}

\newcommand{\pa}{\parallel}

\def\vx{{\bm x}}
\def\vk{{\bm k}}
\def\vka{{\bm k1}}
\def\vkb{{\bm k2}}
\def\vkc{{\bm k3}}
\def\ka{k_{1}}
\def\kb{k_{2}}
\def\kc{k_{3}}
\def\cG{{\cal G}}
\def\fnl{f_{_{\rm NL}}}

\def\pa{{\partial}}
\def\f{\frac}
\def\l{\left}
\def\r{\right}
\def\d{{\rm d}}
\def\cR{\zeta}

\begin{document}


\title{Trans-Planckian Issues for Inflationary Cosmology}

\author{Robert Brandenberger}
\email{rhb@physics.mcgill.ca}
\affiliation{Department of Physics, McGill University, Montr\'eal, QC, H3A 2T8, Canada}
\author{J\'er\^ome Martin}
\email{jmartin@iap.fr}
\affiliation{Institut Astrophysique de Paris, 98bis Boulevard Arago,  
75014 Paris, France}
\affiliation{Research Center for the Early Universe (RESCEU),
Graduate School of Science, The University of Tokyo, Tokyo 113-0033, Japan}


\begin{abstract}

The accelerated expansion of space during the period of
cosmological inflation leads to trans-Planckian issues which
need to be addressed. Most importantly, the physical
wavelength of fluctuations which are studied at the
present time by means of cosmological observations
may well originate with a wavelength smaller than the Planck
length at the beginning of the inflationary phase. Thus,
questions arise as to whether the usual predictions of
inflationary cosmology are robust considering our
ignorance of physics on trans-Planckian scales, and
whether the imprints of Planck-scale physics are at
the present time observable. These and other related
questions are reviewed in this article.

\end{abstract}


\maketitle

\section{Introduction}

The theory of cosmological inflation \cite{Guth} (see
als \cite{Brout, Starob1, Sato} for related original work) has become
the current paradigm of early universe cosmology. Not only does
the scenario solve some of the conceptual problems of Standard
Big Bang cosmology, the previous paradigm of cosmology, but it
provided \cite{Mukh} the first explanation for the origin of 
the large-scale structure of the universe based on causal physics
(see also \cite{Brout, Press, Starob2} for related original work).
Better yet, inflationary cosmology was predictive. The
scenario predicted an almost scale-invariant spectrum of
curvature fluctuations which are coherent and passive on
scales which in the early universe are larger than the
Hubble radius. As has been known since around 1970
\cite{Peebles, SZ}, such a spectrum predicts characteristic
oscillations in the angular power spectrum of cosmic
microwave (CMB) anisotropy maps, oscillations which were
first discovered by the Boomerang experiment \cite{Boomerang}
and later confirmed with high accuracy by the WMAP
satellite experiment \cite{WMAP}. 

Inflationary cosmology is based on the assumption that there
is a period in the very early universe during which space
is expanding at an accelerated rate (typically almost
exponentially). Figure 1 depicts a space-time sketch
of inflationary cosmology. The vertical axis is time,
the horizontal axis represents physical distance. The
period of inflation begins at a time $t_{\rm i}$ and ends at
time $t_{\rm R}$. The accelerated expansion of space during
this time interval can
create a large and almost spatially flat universe from an
initial state in which the size of space is microscopic.
In this way, inflation solves the ``size problem'' of
Standard Cosmology. In an accelerating universe the
contribution of spatial curvature to the total density
decreases. Hence, a spatially flat universe is a local
attractor in initial condition space \cite{Kung} 
\footnote{Note that it is not a global attractor, and hence
a certain degree of spatial flatness is required in
order to be able to enter into a period of accelerated
expansion.}, thus providing a solution to the ``flatness problem''.

In Figure 1 three distance scales are shown.
The solid curve which corresponds to an almost constant
value during the period of inflation is the {\it Hubble
radius} $\ell_{\rm H}(t)$ which is given by the inverse
expansion rate of space. The Hubble radius plays an
important role in the evolution of cosmological perturbations,
the inhomogeneities which give rise to the large-scale structure
of the universe and which lead to CMB anisotropies.
On length scales smaller than the Hubble radius, matter
fluctuations oscillate almost like in flat space-time.
These fluctuations freeze out when the wavelength crosses
the Hubble radius. On larger scales, the evolution of the
perturbations is governed by gravity. The dashed curve
which tracks the Hubble radius before the onset of 
inflation but whose length grows exponentially during
the period of inflation is the {\it horizon}, the forward
light cone from a point at the time of the Big Bang.
The horizon is the largest distance that causal information
can travel. The fact that the horizon becomes exponentially
larger than the Hubble radius provides a solution to the
``horizon (or homogeneity) problem'' of Standard Cosmology: 
In the context of inflation, the causal horizon can be much larger that
any scale which we observe today. In particular this
can explain the near isotropy of the CMB.

The third length scale shown in Figure 1, the curve labelled
by $k$, represents the physical length of a fluctuation
mode. The fact that modes well inside the Hubble radius at
early times become exponentially larger than the Hubble radius
by the end of the period of inflation provides a causal 
mechanism by which fluctuations are generated in inflationary
cosmology. Since classical matter redshifts exponentially
during the period of inflation, it is natural to assume that
matter approaches the vacuum state. The vacuum, however,
is permeated by quantum vacuum perturbations, and
hence it was conjectured \cite{Brout, Press, Starob2} that
the primordial fluctuations are quantum mechanical in nature. The
theory of structure formation in inflationary cosmology is
based on the quantum theory of cosmological perturbations \cite{Mukh2, Sasaki}
(see e.g. \cite{MFB,Martin:2004um} for in-depth reviews, and \cite{RHBfluctrev}
for a pedagogical overview). Based on the above discussion
it is reasonable to assume that fluctuations begin in their quantum vacuum
state \cite{Mukh}. While on sub-Hubble scales, the fluctuation modes
undergo quantum vacuum oscillations. They freeze out once the
wavelength of the mode becomes larger than the Hubble radius. 
Thereafter the modes are squeezed while on super-Hubble scales.
In particular, all modes acquire the same angle in phase space. They
hence enter the Hubble radius at late times as  standing waves,
leading to the characteristic acoustic oscillations in the angular
power spectrum of CMB anisotropies which were mentioned at
the beginning of this section. Based on the time-translation symmetry
of the inflationary phase, it is expected \cite{Press} that the spectrum
of the density perturbations produced during inflation is approximately 
scale-invariant, and the mathematical analysis confirms this
result.

\begin{figure} 
\includegraphics[height=9cm]{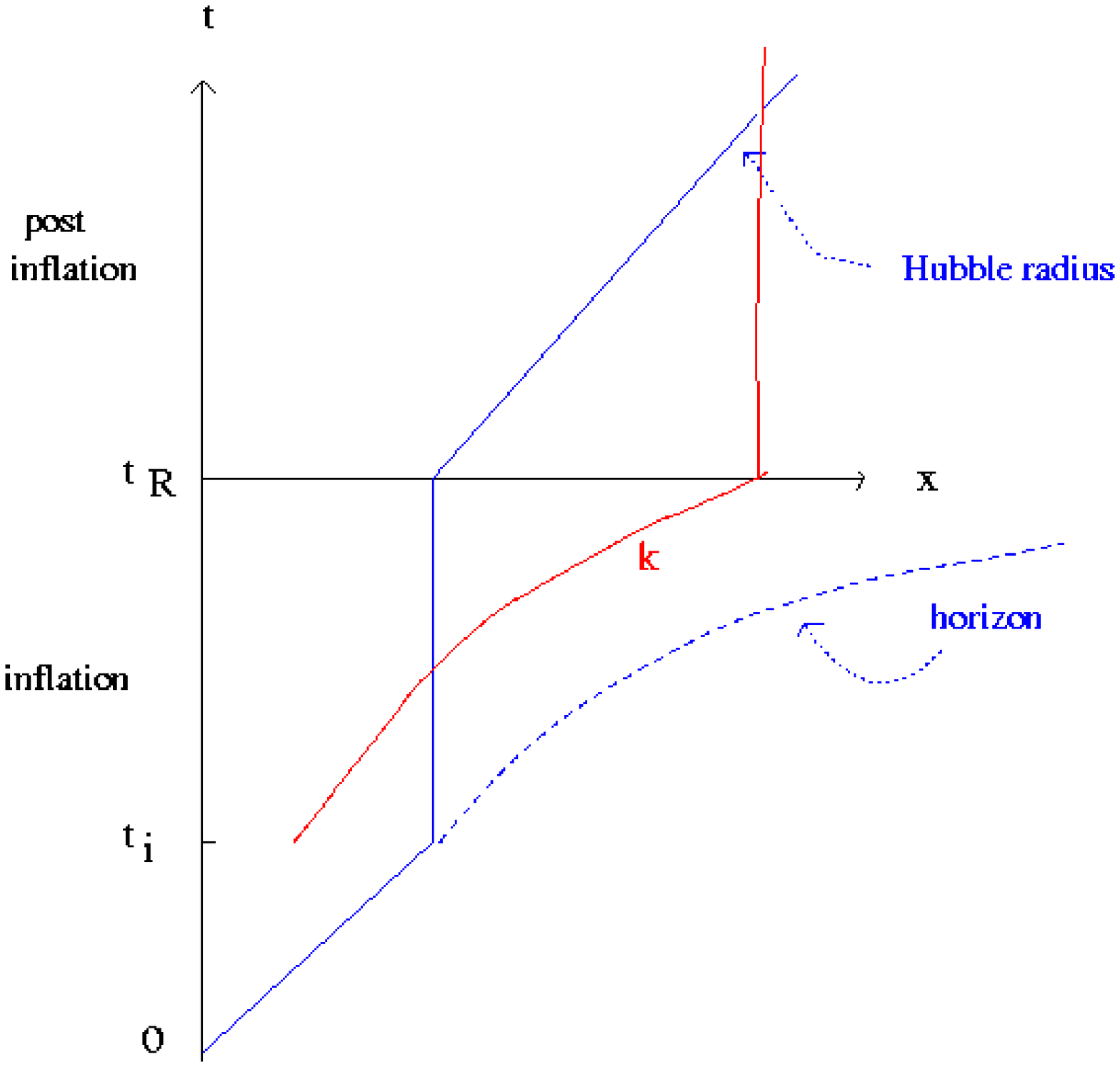}
\caption{Space-time sketch of inflationary cosmology.
The vertical axis is time, the horizontal axis corresponds
to physical distance. The solid line labelled $k$ is the
physical length of a fixed comoving fluctuation scale. The
role of the Hubble radius and the horizon are discussed
in the text.}
\label{infl1}
\end{figure}

The causal generation mechanism of fluctuations described above
is the most significant success of inflationary cosmology. However,
a second look at the space-time sketch of Fig. 1 reveals a serious
conceptual problem \cite{RHBrev0, Jerome}. A sufficiently long
period of inflation is required in order for modes which are
currently probed in cosmological observations to have a wavelength
smaller than the Hubble radius at the beginning of the period of
inflation - which is a necessary condition for the validity of
the inflationary structure formation scenario. However, if the 
period of inflation is only slightly longer 
(70 e-foldings in models in which the energy scale 
at which inflation takes place is close to the scale of Grand Unification), 
then the wavelengths of all fluctuation modes which are currently inside
the Hubble radius were smaller than the Planck length at the
beginning of the period of inflation. We do not understand physics
on length scales smaller than the Planck length. The problem for
inflationary cosmology is that fluctuation modes emerge from
this sub-Planck-wavelength zone of ignorance, as is sketched
in Fig. 2. New physics
is required to truly understand the origin and early evolution of
the fluctuations \footnote{As stressed in \cite{Vikman} in a different
context, for a model of structure formation to be consistent,
it is not sufficient that the energy scale of the background
cosmology is smaller than the Planck scale. If the wavelength of the
fluctuation modes is smaller than the Planck length, the predictions of
the model cannot be trusted.}. 

 \begin{figure}
\includegraphics[height=9cm]{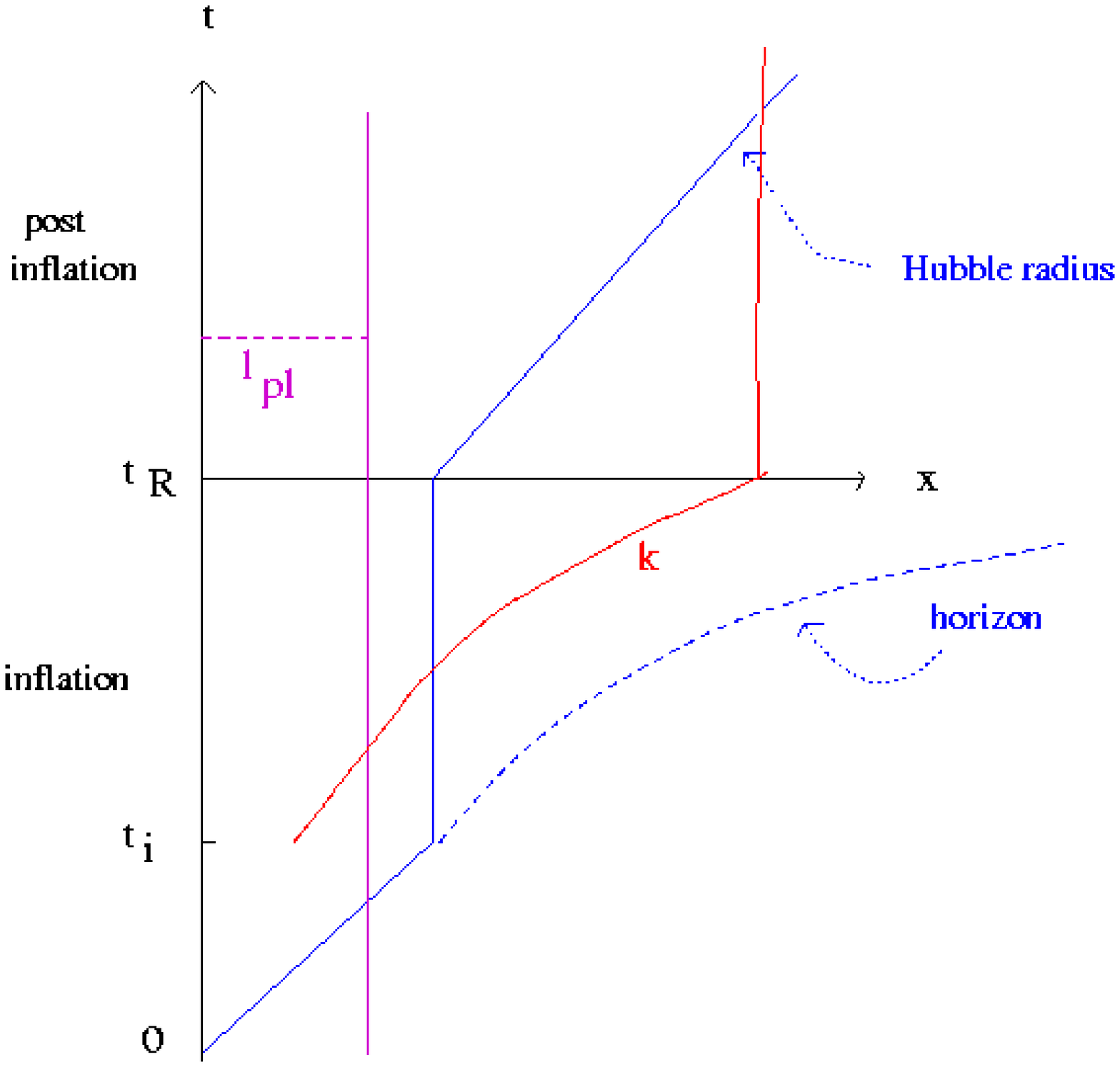}
\caption{Space-time diagram (sketch) 
of inflationary cosmology where we have added an extra
length scale, namely the Planck length $\ell_{_{\rm Pl}}$ 
(majenta vertical line).
The symbols have the same meaning as in Figure 1.
Note, specifically, that - as long as the period of inflation
lasts a couple of e-foldings longer than the minimal value
required for inflation to address the problems of standard
big bang cosmology - all wavelengths of cosmological interest
to us today start out at the beginning of the period of inflation
with a wavelength which is smaller than the Planck length.}
\label{infl2}       
\end{figure}

Based on the geometry of Figure 2, there are good reasons to
expect that Planck-scale physics can lead to a modification of
the spectrum of cosmological perturbations: If we imagine starting
off all perturbation modes on a fixed space-like Cauchy surface
at the beginning of the period of inflation, then short distance
modes will feel the effects of the modified physics for a longer
time than long wavelength modes. If the evolution of the modes
is not adiabatic on short wavelength scales, then one would
expect the spectrum for short wavelength modes to be boosted
relative to that of long wavelength modes. This would produce
a blue tilt in the spectrum of fluctuations.

The above problem is now called the {\it trans-Planckian problem} for
inflationary cosmology. While it may be a problem from the point of
view of inflationary theory, it can instead to viewed as a
{\it trans-Planckian window of opportunity} to probe Planck-scale
physics with current cosmological observations. The accelerated
expansion of space brings scales into the observable range which
we have no chance of probing in accelerator physics.

In the following sections, we will review these trans-Planckian
issues for inflationary cosmology. In Section 2 we discuss
some theoretical approaches, beginning with a brief review
of some of the original approaches to studying the sensitivity
of the predictions of inflationary cosmology to assumptions about
trans-Planckian physics. We continue with a discussion of two
more recent realizations of inflation in the context of specific
models of trans-Planckian physics. In one case we find dramatic
differences compared to the ``standard results'', in the second
case we recover the usual spectrum. We end Section 2 with some
general remarks. In Sections 3 and 4 we discuss observational constraints
on the magnitude of trans-Planckian effects in inflationary
cosmology. Finally, in Section 5 we attempt to draw connections
with some broader issues.

\section{Theoretical Approaches}

\subsection{Original Approaches}

\subsubsection{Modified Dispersion Relations}

The original work on the trans-Planckian problem for inflation
was performed in the context of {\it modified dispersion relations},
\cite{Jerome, Niemeyer} for the linear fluctuation modes, 
following works \cite{Unruh, CJ} which
studied the dependence of black hole radiation on Planck scale
physics. This method represents a toy model to study the
unkown effects of trans-Planckian physics, a method which has
the practical advantage of remaining within the context of applicability of 
linear cosmological perturbation theory. Before introducing
the basics of the method, we must remind the reader of the
relevant formalism.

We are interested in following the {\it scalar metric perturbations},
the linearized gravitational fluctuation modes induced by matter
perturbations. In the context of General Relativity,
The relevant canonical variable (the
Mukhanov-Sasaki variable \cite{Mukh2, Sasaki}) $v$ satisfies
the following Fourier space equation of motion \cite{MFB, RHBfluctrev}
\begin{equation}
\label{pertEoM}
v_{\bm k}'' + \left(k^2 - \frac{z''}{z}\right) v_{\bm k} \, = \, 0 \, ,
\end{equation}
where $k$ denotes the comoving momentum and a prime stands
for the derivative with respect to conformal time $\eta$.
The function $z(\eta)$ depends on the background cosmology.
If the equation of state of the background is time-independent,
then $z(\eta)$ is proportional to the scale factor $a(\eta)$. The
variable $v$ is related to the curvature fluctuation $\zeta$ in
comoving gauge, a coordinate system in which the matter fluctuation
vanishes, via $\zeta = z^{-1} v$. The variable $\zeta$ carries 
vanishing mass dimension.

Of particular interest is the dimensionless power spectrum
${\cal P}_{\zeta}(k)$ defined via [for a more accurate 
definition, see Eq.~(\ref{eq:ngps})]
\be
\label{eq:defpsintro}
{\cal P}_{\zeta}(k) \, \sim \, k^3 |\zeta_{\bm k}|^2 \, .
\ee
The power spectrum is called ``scale-invariant" if ${\cal P}_{\zeta}(k)$
is independent of $k$.  

The idea of the modified dispersion relation method 
is to introduce a non-trivial relation between
the physical frequency and momentum of the modes
\be \label{moddisp}
k^2 \, \rightarrow \, k_{\rm eff}^2(k, \eta) \,
\equiv \, a^2(\eta) \omega_{\rm phys}^2\left[\frac{k}{a(\eta)}\right] \, ,
\ee
where the function $\omega_{\rm phys}(u)$ deviates from
its usual form $\omega_{\rm phys}(u) = u$ only in the
ultraviolet (UV). As will be discussed below, such dispersion relations
arise from specific theories of Planck-scale physics
such as Ho\v{r}ava-Lifshitz gravity \cite{Horava}.

The modification of the dispersion relation implies
a change in the Hamiltonian of the perturbation modes
which has important effects at short wavelengths.
The analysis of~\cite{Jerome} is based on the following
assumptions: first, we consider a space-like hypersurface
on which we set up initial conditions. Second, the initial
conditions are chosen to represent the lowest energy
state of the local Hamiltonian at the initial time.
Since this Hamiltonian will deviate from the usual
one on short wavelength scales, the initial state
will deviate from the usual Bunch-Davies vacuum
of the un-modified theory, the deviations increasing
towards the ultraviolet. 

For mild distortions of the dispersion relation such as
those considered in \cite{Unruh} where $\omega^2$
asymptotes to a constant in the ultraviolet, the evolution
of the mode functions will be adiabatic (they are the
WKB solutions) and track the
local vacuum state. In this case, the evolution will lead to
the usual scale-invariant spectrum at late times.

On the other hand, if the dispersion relation violates
the adiabaticity condition
\be \label{adcond}
\left \vert\frac{3 (\omega')^2}{4 \omega^4} 
- \frac{\omega''}{2 \omega^3} \right \vert \, < \, 1 
\ee
then an initial vacuum state will evolve to a state
which on large scales differs from the Bunch-Davies
vacuum of the unmodified theory. The deviations
increase as the frequency increases, thus leading
to a steep blue spectrum in violation of the observational
constraints (see Sections 3 and 4). 

Deviations from adiabaticity arise for one branch of the
modified dispersion relations proposed in \cite{CJ}.
Specifically, one can consider the example \cite{Lemoine, Jerome2}
\be \label{modexample}
\omega_{\rm phys}^2 \, = \, k_{\rm phys}^2 
- 2 b_{1} k_{\rm phys}^4 + b_{2} k_{\rm phys}^6 \, ,
\ee
with $b_{1}$ and $b_{2}$ being two positive constants. In
the case of this dispersion relation (sketched in Fig. 7)
there can be a region of values of $k_{\rm phys}$ for which the
adiabaticity condition is violated. In this example,
adiabaticity is maintained both in the far ultraviolet and
the infrared. 

While we do not expect concrete models of
modified physics on trans-Planckian scales to yield dispersion
relations which are adiabatic in the far UV, assuming that
this takes place in our toy model makes it easy to justify
our initial conditions: we can start modes in the adiabatic
vacuum in the far UV. As the universe expands, the physical
wavenumber will redshift. Modes which begin in the far UV
adiabatic regime will experience a time interval of finite
duration during which the wavenumber is in the non-adiabatic
region. Hence, the wavefunction will obtain corrections
from its WKB form. In particular, the amplitude of the
final fluctuations will change \cite{Lemoine}. 

To a first approximation (in the deviation of the expansion of space
from exponential), the shape of the spectrum will not change
since all modes spend the same amount of time in the region
of non-adiabaticity (as discussed in Section 3, there will
be superimposed oscillations in the spectrum). 
However, in models in which non-adiabaticty
is violated at all UV scales (like the ones considered in
\cite{Jerome}), short wavelength modes spend more time in the
region of non-adiabaticity, and hence the spectrum of
cosmological perturbations will acquire a blue tilt whose
spectral slope can well exceed current limits.  

\subsubsection{New Physics Hypersurface}

A more conservative approach to the trans-Planckian problem
is simply not to evolve the fluctuation modes during the
time period in which their wavelength is smaller than
the length scale of the new physics. This corresponds
to introducing a time-like {\it new physics hypersurface}
on which initial conditions are imposed. We expect the
wavelength at which the new physics hypersurface is reached
to be given by the Planck length (or the string length - in
the context of string cosmology - if it is larger).

At this point, the trans-Planckian problem has simply been
shifted to the problem of choosing initial conditions on
the new physics hypersurface. The most conservative approach
is to start modes off in their local adiabatic vacuum.
Note, however, that precisely the main point of the analysis
of the previous subsection is that new physics acting
on trans-Planckian scales can well produce states which
are very different from the local adiabatic vacuum at the
time when the wavelength equals the new physics scale.

Even continuing with the conservative approach mentioned
above, a careful tracking of the mode wavefunctions reveals
residual oscillations in the power spectrum 
\cite{Greene, Danielsson, Parentani, Bozza, Schalm}.  
The amplitude of these oscillations depends on whether we are 
computing the spectrum of test scalar fields on a fixed background 
metric, gravitational waves, or scalar metric fluctuations.
In the two former cases the amplitude of the oscillations
in the power spectrum is 
\be
{\cal A}_{\rm GW} \, \sim \, {\cal O}\left(1\right) 
\left( \frac{H}{M_{_{\rm C}}} \right)^2 \, ,
\ee
where $M_{_{\rm C}}$ is the mass scale of the new physics hypersurface,
and in the case of scalar cosmological perturbations we have
\be
{\cal A}_{_{\rm S}} \, \sim \, {\cal O}\left(1\right) 
\frac{H}{M_{_{\rm C}}} \, .
\ee

Note that within the {\it new physics hypersurface approach},
the setting of the initial state is not unique. Whereas
the prescription adopted by \cite{Greene, Danielsson, Schalm}
leads to oscillations of the abovementioned magnitude, the
choice made by \cite{Parentani} and studied in more detail in
\cite{Jerome3} leads to oscillations with amplitude suppressed
by three powers of $H / M_{_{\rm C}}$. The dependence of the amplitude
of the trans-Planckian effects on the definition of the ``local
vacuum state'' was explored in detail in \cite{Bozza}.

In the context of the choice of the vacuum state for cosmological
perturbations there has been some discussion about vacuum
states which are alternatives to the usual Bunch-Davies \cite{BD}
vacuum. These are the {\it $\alpha$ vacua} \cite{alpha}
discussed in the context of cosmological perturbations in
\cite{Danielsson, Lowe}. They are in different superselection
sectors of states than the Bunch-Davies vacuum, and applied
to inflationary cosmology they also lead to oscillations in
the power spectrum of fluctuations. The use of such vacua
has, however, been criticized on the basis of their
instabilities \cite{Shenker2, Larson} (but see \cite{Lowe2}).
Our view, however, is that in the context of inflationary
cosmology which is not past eternal \cite{Borde}, it makes more sense to
restrict attention to states which are in the same superselection
sector as the usual vacuum, and which can hence be generated
from such a vacuum by local trans-Planckian physics.

\subsubsection{Other Approaches}

In the two previous subsections we discussed two general
frameworks for studying trans-Planckian effects on the
spectrum of cosmological perturbations. Another general
framework is the {\it effective field theory} approach.
The idea of effective field theory is to integrate out
the high energy physics and to derive an effective Lagrangian
which only involves observable scales. The application of
effective field theory techniques to the trans-Planckian
problem of inflation was pioneered in \cite{Shenker1}
and \cite{Cliff}. However, the expansion of space (in
particular the fact that Planck-scale modes are redshifted
into the visible sector) leads to challenges which are
not present in usual applications of effective field
theory \footnote{These challenges also arise in some
non-gravitational contexts, e.g. when studying quantum
field theory in strong external electromagnetic fields~\cite{Martin:2007bw}.},
challenges which were not taken into account in \cite{Shenker1},
where it was claimed that trans-Planckian effects
on the spectrum of cosmological perturbations cannot
be larger than ${\cal O}(H / M_{_{\rm C}})^2$. However, as shown e.g.
in the work of \cite{Cliff}, even in the context of
effective field theory and local Lorentz invariance
it is possible for the pre-inflationary dynamics to
produce states which deviate from the standard
Bunch-Davies vacuum and for which hence the corrections
to the power spectrum are larger.
  
Some authors have considered general modifications
of physics on Planck scales such as 
{\it modified uncertainty principles} \cite{Hassan}, 
{\it space-space non-commutativty} \cite{Chu, Greene}
and {\it space-time non-commutativity} \cite{Pei-Ming}
and computed the changes in the spectrum of
perturbations in an inflationary cosmology. All of
these approaches can be viewed as leading to specific
prescriptions of setting the intial conditions on a
space-like new physics hypersurface.
 
If we make the strong assumptions firstly that the microscopic 
structure of space-time on Planck
scales is Lorentz invariant and secondly that the state of the
system is in its local vacuum, then it indeed follows
that the corrections to the spatial correlation function
and hence to the power spectrum are bounded by
${\cal O}(H / M_{_{\rm C}})^2$ \cite{Senatore}. Even if one
accepts the first assumption, the second one is a
very strong one. We know that (in the context of Einstein gravity
with matter fields obeying the usual energy conditions) 
inflation is singular in the past \cite{Borde}, and there
is thus pre-inflationary dynamics which must be taken into
account when determining what the initial state of the
fluctuation modes is. In particular, perfectly Lorentz-invariant
theories can lead to a bouncing universe, e.g. through
the addition of quintom matter \cite{quintombounce} or
in the context of specific string theory backgrounds
\cite{Kounnas}.

\subsection{Specific Models}

\subsubsection{Bouncing Inflation}

In theories which admit non-singular bouncing cosmologies
(e.g. Ho\v{r}ava-Lifshitz gravity in the presence of non-vanishing
spatial curvature \cite{HLbounce}), it is possible to
have a post-bounce inflationary phase and, by time
reflection symmetry, a contracting deflationary phase (see
Figure 3 for the corresponding space-time diagram). The
spectrum of cosmological perturbations was studied
in this context in \cite{Xinmin}. 

\begin{figure}[htbp] 
\includegraphics[height=9cm]{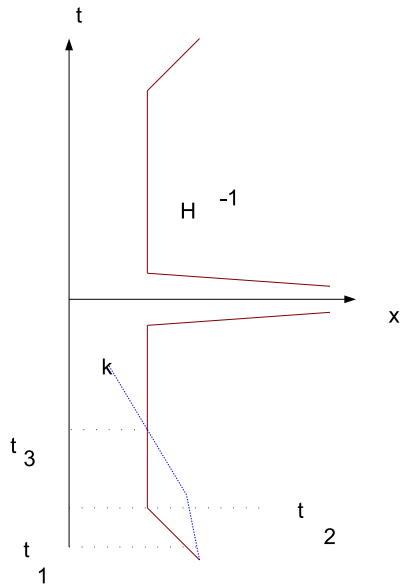}
\caption{Space-time sketch of the bouncing inflation model. The vertical axis
is time, the horizontal axis denotes the physical space coordinate.
The Hubble radius ${H}^{-1}$ is constant during the deflationary phase
of contraction and during the inflationary phase of expansion. It diverges
around the bounce point. The solid (blue) curve labelled $k$ indicates
the wavelength of a fluctuation mode which exits the Hubble radius in its
Bunch-Davies vacuum state during the radiation phase of contraction at
time $t_1(k)$, evolves on super-Hubble scales until it enters the Hubble
radius at time $t_3(k) = - t_i(k)$ during the deflationary phase. The time $t_2$
is the transition between radiation contraction and deflation. }
\label{bounce2}
\end{figure}

Since the universe begins large and cold, it is 
natural to start with perturbations in their Bunch-Davies
vacuum state. In a contracting universe the curvature
fluctuation variable $\zeta$ grows on super-Hubble
scales (see e.g. \cite{Wands, FB}). In a radiation phase of contraction the
growth rate is proportional to $\eta^{-1}$ (the conformal
time $\eta$ is increasing towards $0$), and in
the deflationary phase $\zeta$ increases as $\eta^{3}$
($\eta$ is increasing to $\infty$ in this phase). In
this process, the initial vacuum spectrum is transformed
into a dimensionless spectrum which scales as
\be \label{spectrum}
{\cal P}\left[k, - t_i\left(k\right)\right] \, \sim \, \, k^{-4} \, ,
\ee
where $- t_i(k)$ is the time when the scale $k$ enters
the Hubble radius during the deflationary phase. 

The evolution of the fluctuations between the time $t_i(k)$
of Hubble radius entry during the contracting phase and
Hubble radius exit at the time $t_i(k)$ is symmetric. Hence,
the spectrum at Hubble radius exit is the same as at
Hubble radius entrance. Since $\zeta$ is conserved on
super-Hubble scales in the expanding phase, the
final spectrum of curvature fluctuations in the expanding
phase is given by (\ref{spectrum}), i.e. it is an $n_s = -3$ spectrum. 
This is a highly red spectrum
compared to a scale-invariant one ($n_{_{\rm S}} = 1$ which is
indicated by experiments). The redness of the spectrum
is due to the extra growth which long wavelength
modes undergo since they spend a longer time on
super-Hubble scales.

As will be discussed in the following section, a red
spectrum is not only constrained by observations,
but also by back-reaction. Applied to our model,
this implies that the bouncing model cannot be
symmetric about the bounce point. The phase of
deflation must be short. As a consequence, one obtains
a scale-invariant spectrum in the ultraviolet and a
red spectrum in the infrared.

This model is an example of how trans-Planckian
effects can lead to very large deviations from
scale-invariance. As in most non-singular bouncing
cosmologies, it requires specially tuned matter
and initial condition modeling to obtain a scenario
in which a period of deflation follows after a radiation
phase. One way, as explored in the quintom
matter bounce scenarios \cite{quintombounce}
is to have the pre-bounce radiation come from
a scalar field $\varphi$ with quartic potential. While
the scalar field is oscillating about its ground state,
the time-averaged equation of state is that of
radiation. In the contracting phase the amplitude
of oscillations increases. But once the amplitude
exceeds the Planck scale, a deflationary slow-climb
phase sets in.

\subsubsection{Ho\v{r}ava-Lifshitz Inflation}

Ho\v{r}ava-Lifshitz (HL) gravity is a four space-time-dimensional
theory which has been proposed \cite{Horava} as a
power-counting renormalizable theory of quantum gravity.
In this theory, the microscopic Lagrangian is not
locally Lorentz-invariant. There is a distinguished time
direction, and the gravitational Lagrangian contains
higher space-derivative terms. As a consequence,
space-time diffeomorphism invariance is also lost.
The theory is only invariant under the subgroup of
spatial diffeomorphisms, and under separate
space-independent time reparametrizations. The guiding principle
of the construction of the Lagrangian is invariance 
with respect to the residual symmetries and
power-counting renormalizability with respect to
an anisotropic scaling
\be
{\bm x} \rightarrow b {\bm x} \,\,\, , \,\,\, t \rightarrow b^3 t \, ,
\ee
where $b$ is some constant. There are two versions
of the theory - the {\it projectable} version in which
the lapse function $N$ \footnote{In the Hamiltonian approach
to gravity, the lapse function is the fluctuation in the time-time
component of the metric.} is a function of only time, and
the {\it non-projectable} version in which $N$ can
depend on space and time.

Renormalizability allows up to six-space-derivative terms in the
equations of motion. Hence, quite naturally a modified
dispersion relation for fluctuation modes results. There is
a second major difference in the theory of cosmological
perturbations in HL gravity compared to Einstein gravity:
because of the reduced symmetry, there is an
extra scalar gravitational fluctuation mode (see e.g.
\cite{HLreview} for a review of HL gravity). It has been
shown \cite{Cerioni1} that in the case of the projectable
version of HL gravity the extra scalar mode is sick (either
ghost-like or tachyonic), while in the non-projectable
version it can be \cite{Cerioni2} well-behaved (for suitable
choices of the free parameters of the Lagrangian). The
extra mode in fact becomes massive in the infrared and thus
only plays an important role in the ultraviolet.

Because of the existence of the extra degree of freedom
and because of the modified dispersion relation one
might expect that the spectrum of cosmological perturbations
in inflationary cosmology in the context of HL gravity
would lead to a very different spectrum of cosmological
perturbations than in Einstein gravity. However, a
careful analysis of this issue \cite{Elisa} has shown
firstly that the evolution of the regular fluctuation mode is
adiabatic and secondly that the fluctuations induced
by the extra degree of freedom are sub-dominant.
Hence, up to oscillations in the spectrum whose frequency
is too high to be observationally detectable, the
spectrum of cosmological perturbations which emerges
when starting all modes off at a fixed time in their instantaneous
minimum energy state is scale invariant.

This example indicates a certain degree of
robustness of the standard predictions of inflation even
if the underlying physics at the Planck scale is substantially
modified.

\subsection{General Comments}

It is interesting to ask to which extend the trans-Planckian
problem discussed here is specific to inflationary cosmology
or what aspects of the problem arise in any cosmological
background. To address this question, let us introduce two
paradigms alternative to inflation which can explain the
observed almost scale-invariant spectrum of cosmological
fluctuations. The first is the {\it string gas cosmology}
model introduced as a background cosmology in \cite{BV} (see
also \cite{Perlt}) and shown to lead to a mechanism for
producing an almost scale-invariant spectrum of cosmological
perturbations in \cite{NBV, BNPV1}. The second is the 
{\it matter bounce} scenario of \cite{Wands, FB}, a non-singular
bouncing cosmology with an initial matter-dominated phase of
contraction.

{\it String gas cosmology} (see e.g. \cite{SGCrev} for a
recent review) is based on the realization that
a gas of closed strings has a maximal temperature, the
``Hagedorn temperature'' \cite{Hagedorn} $T_{\rm H}$. At temperatures
close to $T_{\rm H}$, all of the modes of strings including
winding modes are excited. This high temperature phase is
called the ``Hagedorn phase''. It is then reasonable to 
assume that the universe begins in the Hagedorn phase.
Since the temperature of a box of strings is independent
of the radius $R$ of the box for a wide range of values of $R$
(assuming that the entropy of the system is large), it is
not unreasonable to assume that the Hagedorn phase is
quasi-static (including fixed dilaton). Eventually (through
the decay of string winding modes into string loops)
the universe transits into the radiation phase of Standard
Cosmology. The time when this happens is called $t_{\rm R}$ in
analogy to the reheating time at the end of inflation. In the
context of this string gas cosmology background,
it was realized in \cite{NBV, BNPV1} that thermal string
fluctuations in the Hagedorn phase induce a scale-invariant
spectrum of curvature fluctuations at late times with a slight
red tilt, and a scale-invariant spectrum of gravitational
waves with a slight blue tilt \cite{BNPV2} - a key prediction
of string gas cosmology with which it can be observationally
distinguished from inflationary cosmology.

\begin{figure} 
\includegraphics[height=10cm]{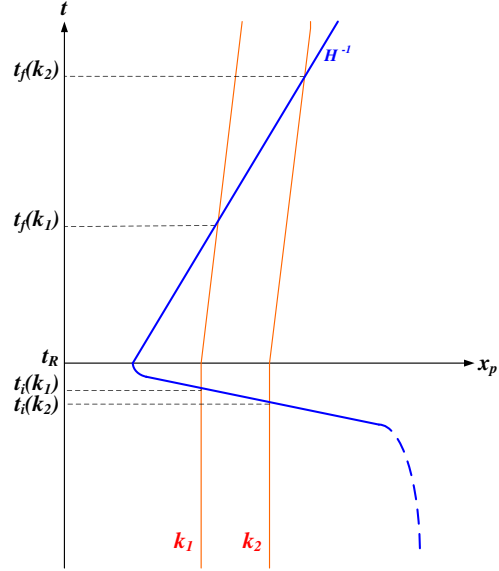}
\caption{Space-time diagram (sketch) showing the evolution of fixed 
co-moving scales in string gas cosmology. The vertical axis is time, 
the horizontal axis is physical distance.  
The solid curve represents the Einstein frame Hubble radius 
$H^{-1}$ which shrinks abruptly to a micro-physical scale at $t_{\rm R}$ and then 
increases linearly in time for $t > t_{\rm R}$. Fixed co-moving scales (the 
dotted lines labeled by $k_1$ and $k_2$) which are currently probed 
in cosmological observations have wavelengths which are smaller than 
the Hubble radius before $t_{\rm R}$. They exit the Hubble 
radius at times $t_{\rm i}(k)$ just prior to $t_{\rm R}$, and propagate with a 
wavelength larger than the Hubble radius until they reenter the 
Hubble radius at times $t_{\rm f}(k)$.}
\label{spacetimenew}
\end{figure}

Figure 4 shows a space-time sketch of string gas cosmology.
If the Hagedorn temperature is taken to be similar to the
post-inflation temperature (which is typically comparable
to the scale of Grand Unification), then the post-reheating
evolution in inflationary cosmology is identical to the
post-Hagedorn dynamics in string gas cosmology. Working
backwards from the present time, it is not hard to see that
the physical wavelengths of fluctuations which are observed
today are many orders of magnitude larger than the Planck
scale at the time $t_{\rm R}$ (1 mm is a typical value).

Since in string gas cosmology scales which are observed
now never had a wavelength close to the Planck scale,
the trans-Planckian problem for cosmological perturbations
is absent. The basic problems of quantum field theory
in an expanding universe (e.g. time-dependence of the
Hilbert space of Fourier modes \cite{Weiss}) still are
present, but they are not directly coupled to observations,
unlike in inflationary cosmology where they are.

\begin{figure}[htbp] 
\includegraphics[height=9cm]{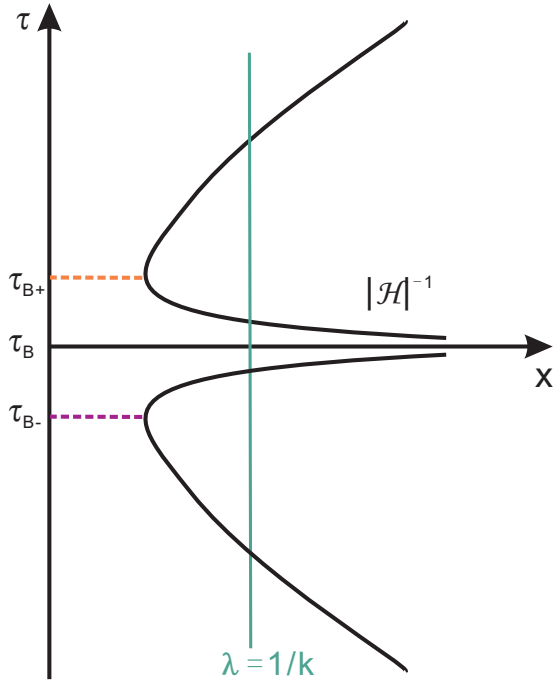}
\caption{Space-time sketch in the matter bounce scenario. The vertical axis
is conformal time $\eta$, the horizontal axis denotes a co-moving space coordinate.
Also, ${\cal H}^{-1}$ denotes the co-moving Hubble radius.}
\label{bounce}
\end{figure}

Figure 5 represents a space-time sketch of a {\it matter bounce}
cosmology. The simplest way to visualize the background 
space-time is to take a mirror inverse of our expanding
Standard Cosmology space-time (which is a contracting
universe which begins in a phase of matter-domination)
and match it to the expanding phases of Standard Cosmology
via a short non-singular bounce phase (see \cite{RHBrev5}
for a recent review of the matter bounce scenario). It
goes without saying that new physics (either in the matter
or in the gravitational sector) is required to obtain such
a non-singular bounce. If the energy scale of the non-singular
bounce is comparable to the scale of Grand Unification, then
- as is the case in string gas cosmology - the physical
wavelengths of scales which are currently probed in
cosmological observations never enter the sub-Planck length
zone of ignorance, and hence there is no trans-Planckian problem
for cosmological perturbations.

As first realized in \cite{Wands, FB}, initial vacuum fluctuations
which exit the Hubble radius during the matter-dominated phase
of contraction acquire a scale-invariant spectrum of curvature
perturbations on super-Hubble scales. The key point is that
curvature perturbations in a contracting universe grow on
super-Hubble scales, and in a matter-dominated background
the extra growth which long wavelength modes experience
because they are super-Hubble for a longer time is exactly
the right amount to convert a blue vacuum spectrum into
a scale-invariant one. One of the key predictions of the
matter bounce scenario is that there is a cutoff scale set
by the transition time between matter domination and radiation
domination during the contracting period below which the
spectrum of perturbations turns blue \cite{LiHong}. In
addition, the growth of the curvature fluctuation on super-Hubble
scales leads to a particular shape and reasonably large 
amplitude of the bispectrum \cite{bispectrum}.


\section{Theoretical Framework}

We have seen in the last sections that trans-Planckian physics 
could modify the inflationary power spectrum. Since we measure the 
inflationary power spectrum when we measure the CMB anisotropies, this 
opens the possibility to test these effects with astrophysical data. 
In order to carry out this program, we must first understand how to 
calculate the trans-Planckian corrections. At this point, we meet a first 
difficulty. Since the physics that controls the shape of these corrections 
is by definition unknown, it seems impossible to establish a generic result 
with regards to the shape of the modified power spectrum. Here we will
consider a conservative starting point, namely that the perturbations
begin in the expanding phase in their instantaneous minimum energy
state (as we have seen before, this is not the most general case).

We have seen before that, if trans-Planckian physics is adiabatic, then 
the inflationary initial conditions are not modified. This means that 
the sub-Hubble Mukhanov-Sasaki variable is still given by
\begin{equation}
\label{eq:vsubhubble}
v_{\bm k}(\eta)\simeq \frac{\alpha_{\bm k}} {\sqrt{2k}} \frac{4\sqrt{\pi}}{\mpl} {\rm e}^{ik\eta}
+ \frac{\beta_{\bm k}}{\sqrt{2k}} \frac{4\sqrt{\pi}}{\mpl}  {\rm e}^{-ik\eta}  
\end{equation}
with $\alpha_{\bm k}=1$ and $\beta_{\bm k}=0$. If, however, 
the physical conditions that prevailed in the trans-Planckian regime were 
non-adiabatic, then there was particle production and, as a result, 
$\beta_{\bm k}\neq 0$. This immediately implies that the power spectrum 
(which is proportional to the square modulus of $v_{\bm k}$) contains 
super-imposed oscillations. Therefore, even without knowing in detail 
the trans-Planckian physics, it is possible to establish a generic prediction 
for the shape of the modified power spectrum. This generic prediction was 
made for the first time in Refs.~\cite{Jerome}.

However, clearly, if we want to make more precise predictions and, for 
instance, compute the amplitude, the frequency or the phase of those 
oscillations, we need to make further assumptions. The most general 
approach consists in parameterizing the deviations from the standard 
situation in the initial conditions. Let $M_{_{\rm C}}$ be the energy scale 
above which some new physics is operating. This scale could be the Planck 
scale or the string scale for instance. A Fourier mode emerges from the 
regime where the new effects are relevant when its physical wavelength 
equals the new length scale introduced in the problem, namely
\begin{equation}
\label{eq:definitialtime}
\lambda(\eta)=\frac{2\pi}{k}a(\eta)=\ell_{_{\rm C}}
\equiv \frac{2\pi}{M_{_{\rm C}}},
\end{equation}
where $k$ is the comoving wavenumber. It is important to notice 
that one can have $\lambda\ll \ell_{_{\rm C}}$ and, at the same time, 
$H\ll \mpl $ where $H$ is the Hubble parameter during inflation. In 
other words, the Fourier mode wavelength can be much smaller than the 
new fundamental scale in a regime where spacetime can still be described 
by a classical FLRW background. The initial conformal time satisfying
Eq.~(\ref{eq:definitialtime}) depends on the scale $k$ and, for this reason, 
it will be denoted in the following by $\eta _{\bm k}$. At this time, if the 
trans-Planckian regime was non-adiabatic, then we expect the initial 
conditions to be modified compared to the standard case. Technically, 
this is expressed by 
\begin{eqnarray}
v_{\bm k}\left(\eta _{\bm k}\right) &=& \frac{\alpha_{\bm k}+\beta_{\bm k}}
{\sqrt{2\omega\left(\eta _{\bm k}\right)}}
\frac{4\sqrt{\pi}}{\mpl},
\\
v_{\bm k}'\left(\eta _{\bm k}\right) &=&
\sqrt{\frac{\omega\left(\eta _{\bm k}\right)}{2}}
\frac{4\sqrt{\pi}\left(\alpha_{\bm k}-\beta_{\bm k}\right)}
{\mpl},
\end{eqnarray}
where $\alpha_{\bm k}$ and $\beta_{\bm k}$ are two complex numbers 
(already introduced before) satisfying 
$\vert \alpha_{\bm k}\vert^2-\vert \beta_{\bm k}\vert^2=1$. These numbers 
completely characterize the influence of the new physics on the initial 
conditions. Again, a priori, a complete calculation of $\alpha_{\bm k}$ 
and $\beta_{\bm k}$ requires the knowledge of the physics beyond the 
scale $M_{_{\rm C}}$. However, if the ratio $H/M_{_{\rm C}}$ goes to zero, 
then we expect to recover the standard Bunch-Davies situation for which 
$\alpha_{\bm k}=1$ and $\beta_{\bm k}=0$. Therefore, if the expressions of 
$\alpha_{\bm k}$ and $\beta_{\bm k}$ are perturbative in $H/M_{_{\rm C}}$, 
then, without loss of generality, one can write
\begin{eqnarray}
\label{eq:alpha}
\alpha_{\bm k} &=& 1+y\frac{H}{M_{_{\rm C}}}+{\cal O}\left(\frac{H^2}{M_{_{\rm C}}^2}\right), \\
\label{eq:beta}
\beta_{\bm k} &=& x\frac{H}{M_{_{\rm C}}}+{\cal O}
\left(\frac{H^2}{M_{_{\rm C}}^2}\right),
\end{eqnarray}
where $x$ and $y$ are two numbers characterizing the perturbative expansion. 
The modified power spectrum can be then determined perturbatively in terms 
of the free parameters $x$ and $y$. Notice that, strictly speaking, $x$ and 
$y$ can be scale-dependent. Here, for simplicity, and in order not to 
introduce arbitrary functions in the problem, we assume that they are roughly 
scale-independent over the range of wavenumbers relevant to the CMB. If 
this is not the case, then one looses the ability to parameterize the 
modified power spectrum in a simple way and one would really need a complete 
theory of the trans-Planckian effects to establish the shape of the 
corrections. It is also important to recall that $\alpha_{\bm k}$ and 
$\beta_{\bm k}$ must satisfy the Wronskian condition. At leading order 
in $H/M_{_{\rm C}}$, this simply amounts to $y+y^*=0$.

We are now in a position where the modified power spectrum can be derived. 
Following Ref.~\cite{Jerome3}, it can be written as (of course, there is a similar 
formula for tensor perturbations that we do not show here)
\begin{eqnarray}
\label{eq:powerspectrum}
k^3P_{\zeta} &=& \frac{H^2}{\pi \epsilon_1\mpl^2}
\Biggl\{1-2\left(C+1\right)\epsilon_1-C\epsilon_2
-\left(2\epsilon_1+\epsilon_2\right)\ln \frac{k}{k_*} 
-2\vert x\vert \frac{H}{M_{_{\rm C}}}
\biggl[1-2(C+1)\epsilon_1-C\epsilon_2
\nonumber \\ & &
-\left(2\epsilon_1+\epsilon_2\right)\ln \frac{k}{k_*}\biggr]
\cos \left[\frac{2M_{_{\rm C}}}{H}
\left(1+\epsilon_1+\epsilon _1\ln \frac{k}{a_0M_{_{\rm C}}}
\right)+\varphi\right] \\ & &
-\vert x\vert \frac{H}{M_{_{\rm C}}}\pi 
\left(2\epsilon_1+\epsilon_2\right)
\times
\sin \left[\frac{2M_{_{\rm C}}}{H}
\left(1+\epsilon_1+\epsilon_1\ln \frac{k}{a_0M_{_{\rm C}}}
\right)+\varphi\right]\Biggr\}. \nonumber
\end{eqnarray}
In this expression, $\epsilon_1$ and $\epsilon_2$ are the two first 
slow-roll, or horizon-flow, parameters. The scale $k_*$ is the pivot 
scale and $a_0$ is the scale factor evaluated at a time $\eta _0$ during 
inflation (the Hubble parameter which appears in the overall normalization 
is also evaluated at the same time). This time is arbitrary and does not 
depend on the scale $k$. For convenience, we choose $k_*/a_0=M_{_{\rm C}}$. 
The quantity $C$ is given by $C\equiv \gamma _{_{\rm E}}+\ln 2-2$ where 
$\gamma _{_{\rm E}}$ is the Euler constant. Finally, $\varphi$ is the phase 
of the complex number $x$, that is to say 
$x\equiv \vert x\vert {\rm e}^{i\varphi}$. It is interesting to notice that, 
at this order, the number $y$ does not appear in the 
expression for $k^3P_{\zeta}$.

\par

\begin{figure}
\includegraphics[width=1.\textwidth,height=.65\textwidth]{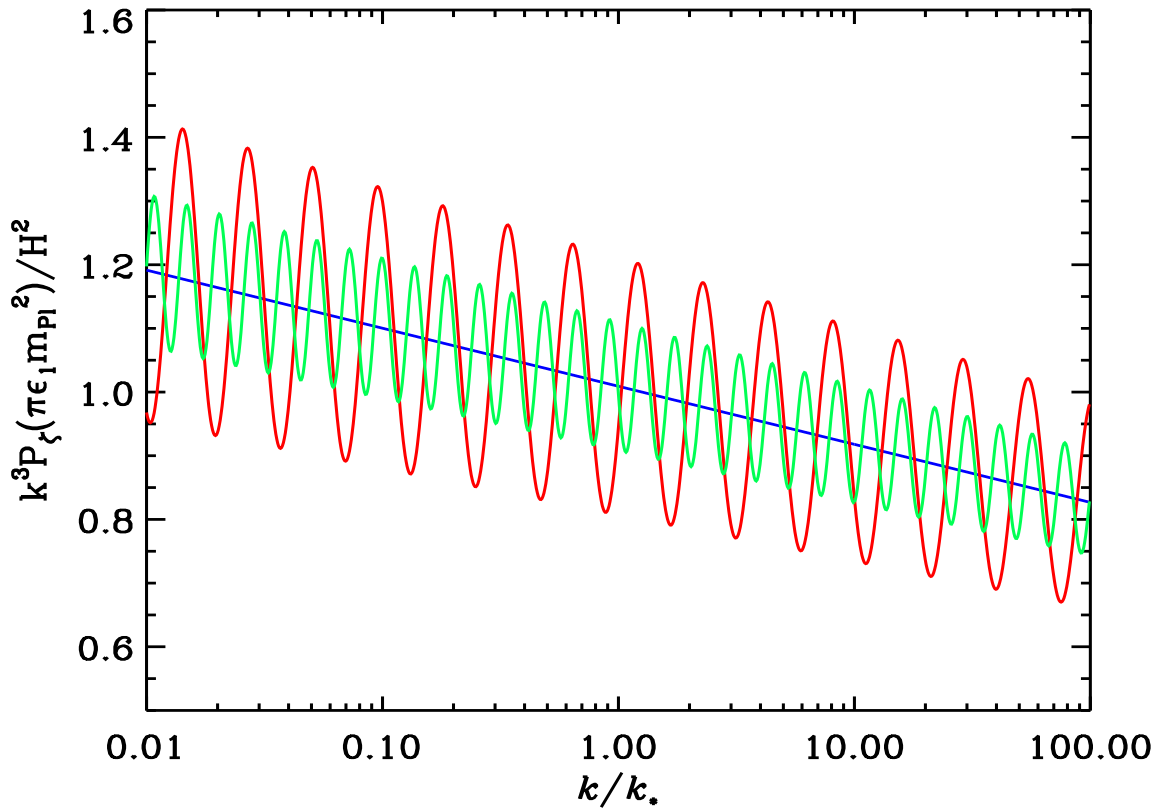}
\caption{Trans-Planckian power spectra given by Eq.~(\ref{eq:powerspectrum}). 
The blue line corresponds to a vanilla model with $\vert x\vert =0$ and 
$\epsilon_1=1/(2\Delta N_*)$, $\epsilon_2=1/\Delta N_*$ with 
$\Delta N_*\simeq 50$ as predicted for the $m^2\phi^2$ inflationary model. 
The red line corresponds to a model with the same values for the slow-roll 
parameters and $H/M_{_{\rm C}}\simeq 0.002$, $\vert x\vert \simeq 50$, $\psi=3$. 
Finally, the green line represents a model with $H/M_{_{\rm C}}\simeq 0.001$, 
$\psi=2$ and the same values for the other parameters.}
\label{fig:powerspectrum}
\end{figure}

Let us now comment on the power spectrum itself. We see it has the 
expected shape. There is the usual slow-roll (almost scale-invariant) part, 
with the overall normalization given by the square of the Hubble parameter 
(during inflation), measured in Planck units, and divided by the first 
slow-roll parameter. This slow-roll part is corrected by two oscillatory 
terms which represent the super-imposed oscillations. It is interesting to 
notice that these oscillations are logarithmic in $k$. This is due to 
the fact that the initial conditions have been chosen on the 
surface $M_{_{\rm C}}=\mbox{constant}$~\cite{Chen:2011zf,Chen:2012ja}. If, on
the contrary, the initial conditions were chosen on a surface of constant 
time, then the oscillations would be linear. This is for instance what happens in 
bouncing models. The amplitude of the oscillations is, 
roughly speaking (taking $|x| \sim 1$), given by $\vert x\vert H/M_{_{\rm C}}$ 
while the frequency is proportional to $(H/M_{_{\rm C}})^{-1}$. Therefore, if 
the initial state is the instantaneous Minkowski vacuum, the amplitude is 
just inversely proportional to the frequency. This means that high 
frequency modes necessarily have a small amplitude. But, of course, there 
is a priori no reason to assume that the frequency and the amplitude are 
fully correlated. In addition, as we are going to see, postulating 
$\vert x\vert=1$ would prevent us to find the best fit.

On general grounds, we expect the ratio $H/M_{_{\rm C}}$ to be a small number. 
Indeed, we know from the COBE normalization that $H\lesssim 10^{-5}\mpl$. We 
do not know the scale $M_{_{\rm C}}$ but reasonable candidates are the Planck 
scale itself or the string scale $M_{_{\rm S}} \simeq 10^{-1}-10^{-3} \, \mpl$. 
This means that $H/M_{_{\rm C}}$ is at most of order $\sim 0.01$. Therefore, 
if $\vert x\vert =1$, the amplitude of the trans-Planckian corrections is 
very small and it will be difficult to detect an effect in the CMB, even
with an experiment like the Planck satellite. 

On the other hand, if $\vert x\vert \neq 1$, one can easily reach a 
level which would be detectable. In this case, we nevertheless meet 
another issue, namely the backreaction problem \cite{Starob, Tanaka}. 
If $\vert x\vert \neq 1$, we start from a non-vacuum state, a natural 
assumption if physics beyond the scale $M_{_{\rm C}}$ is non-adiabatic. 
But a non-vacuum state means that particles are initially present. Since 
these particles carry energy density, there is now the danger that this 
energy density be comparable to or larger than that of the background, 
(which is $\sim H^2\mpl^2$), in which case the energy in the particles
would prevent the onset of inflation. Clearly, the larger $\vert x\vert$ is, 
the larger the trans-Planckian corrections become, and the more severe is the 
danger to have a backreaction problem. In fact, it is easy to estimate the 
regime where there is no backreaction problem. The energy density carried by 
the "trans-Planckian particles" is 
\begin{equation}
\rho_{_{\rm UV}} \simeq M_{_{\rm C}}^4\vert \beta_{\bm k}\vert ^2 \, 
\simeq \, M_{_{\rm C}}^2H^2\vert x\vert ^2. 
\end{equation}
Therefore, the condition $\rho_{_{\rm UV}}\lesssim H^2\mpl^2$ reduces to 
$\vert x\vert \lesssim \mpl/M_{_{\rm C}}$ or, using the COBE 
normalization $H\simeq \mpl 10^{-4}\sqrt{\epsilon_1}$, 
\begin{equation}
\label{eq:backreaction}
\vert x\vert \frac{H}{M_{_{\rm C}}}\lesssim \frac{10^4}{\sqrt{\epsilon_1}}
\left(\frac{H}{M_{_{\rm C}}}\right)^2.
\end{equation}
As a consequence, one can have $\vert x\vert H/M_{_{\rm C}}$ of order one and 
easily satisfy the above inequality (in particular, in inflationary models 
where the gravity waves contribution is extremely small). Of course, this 
does not mean that one should see trans-Planckian effects in the CMB. It 
could very well turn out that $\vert x\vert $ is not large enough to 
compensate the smallness of the ratio $H/M_{_{\rm C}}$. In that case, the 
trans-Planckian effects are there but it seems difficult to imagine how 
the small oscillations could be measured one day.

\begin{figure*}
\includegraphics[width=1.\textwidth,height=0.65\textwidth]{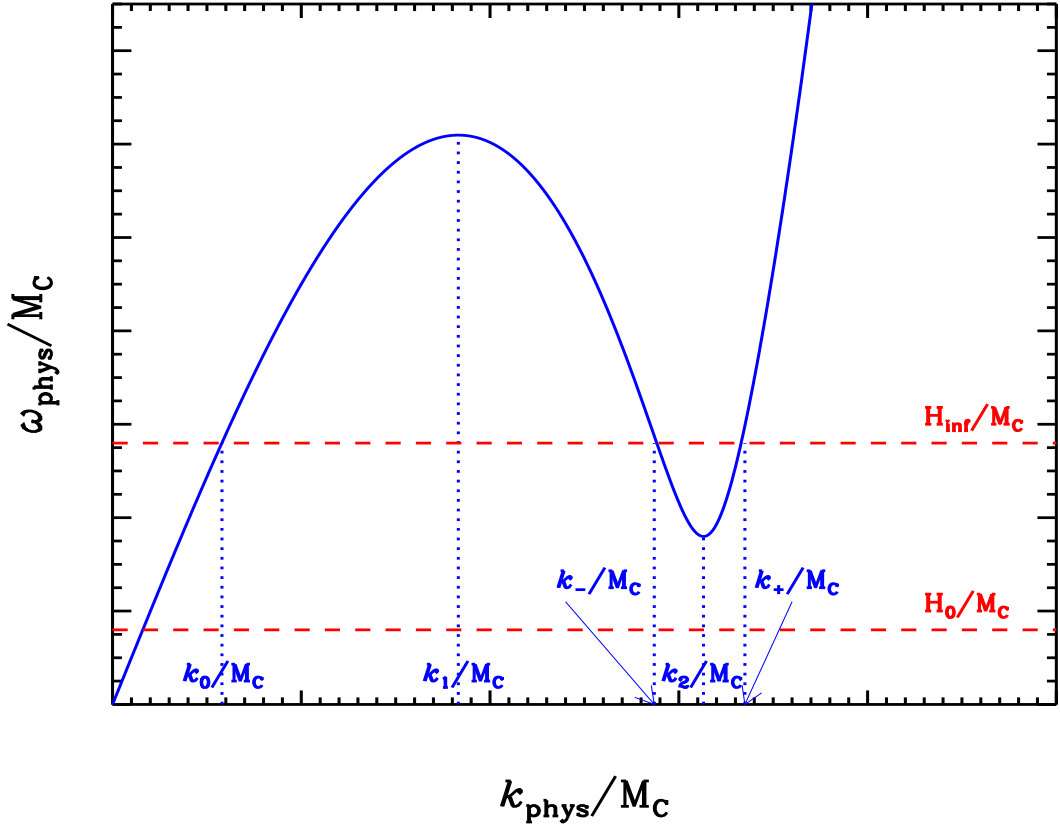}
\caption{Example of a non-linear dispersion relation leading to a modified 
power spectrum. In the regime $k_{\rm phys}\ll M_{_{\rm C}}$, the dispersion 
relation is approximatively linear, as appropriate for standard massless 
excitations, while in the regime $k_{\rm phys}\gg M_{_{\rm C}}$, it strongly 
deviates from linearity due to trans-Planckian effects. In the regions 
where $w_{\rm phys}<H$, the adiabatic approximation is violated and 
trans-Planckian particles production occurs. The horizontal dashed red 
lines represent the value of the Hubble parameter at different times 
during the cosmic evolution. In this model, particles production happens 
during inflation for modes such that 
$k_{\rm phys}<k_0$ and $k_{\rm phys}\in [k_-,k_+]$ and, after inflation, only 
for modes $k_{\rm phys}<k_0$, this latter regime taking place in the linear 
part of the dispersion relation and corresponding to the standard 
amplification of cosmological perturbations on super-Hubble scales.}  
\label{fig:disp}
\end{figure*}

Another remark is in order here. In the above calculation, we have used 
the value of the Hubble parameter during inflation. As noticed in 
Ref.~\cite{Brandenberger:2004kx}, if instead one had used its present day value, the 
no-backreaction condition would have become almost impossible to 
satisfy. Using the Hubble parameter during inflation assumes that 
the trans-Planckian production of particles stops after the end of 
inflation. It is easy to find a situation where this happens. Let us 
for instance consider the case where the trans-Planckian corrections 
arise from a modified dispersion relation (see the previous sections). 
A typical situation is represented in Fig.~\ref{fig:disp} (the model and the 
argument presented here were also studied in Ref.~\cite{Danielsson:2004xw}). 
The dispersion 
relation is almost linear for $k_{\rm phys}/m_{_{\rm C}}\ll 1$ as it should 
be for obvious phenomenological reasons, and deviates from linearity for 
$k_{\rm phys}/m_{_{\rm C}}\gg 1$. In the regimes where 
$\omega_{\rm phys}\lesssim H$, the adiabatic approximation is violated and 
particle production takes place. This is of course the case in the regime 
$k_{\rm phys}\lesssim k_0$ during inflation, this situation corresponding 
to the case where the wavelength of a Fourier mode is larger than the 
Hubble radius. But, as can be noticed in Fig.~\ref{fig:disp}, it is also 
the case when $k_{\rm phys}\in [k_-,k_+]$ during inflation. In this regime, 
adiabaticity is violated and the resulting inflationary power spectrum is 
modified (and, typically, as argued before, acquires super-imposed 
oscillations). But the Hubble parameter evolves (decreases) during inflation 
and after. As a consequence, the horizontal red dashed line in
Fig.~\ref{fig:disp} representing $H/M_{_{\rm C}}$ goes down in this diagram. 
At some point, it passes below the minimum of the dispersion relation and 
particle production stops. In particular, this is the case for the line 
representing the current Hubble parameter. In such a situation, it is clear 
that $H_{\rm inf}$ (and not $H_0$) should be used in order to evaluate 
$\rho_{_{\rm UV}}$.

Moreover, we want $\rho_{_{\rm UV}}\ll H^2\mpl^2$ because the energy 
density of the "trans-Planckian particles" could spoil inflation. But, of 
course, this rests on the prejudice that the corresponding equation of 
state is different from $p/\rho=-1$, typically that of radiation $p/\rho=1/3$. 
When the dispersion relation is modified, as shown in Refs.~\cite{Brandenberger:2004kx}, the 
equation of state also acquires corrections. And, as was demonstrated in
Ref.~\cite{Brandenberger:2004kx}, in the case of the dispersion relation in 
Fig.~\ref{fig:disp}, the modified equation of state is precisely very 
close to that of the vacuum. Therefore, even if one has a backreaction 
problem, it is not obvious that this will prevent inflation to proceed.

\begin{figure*}
\includegraphics[width=0.9\textwidth,height=.55\textwidth]{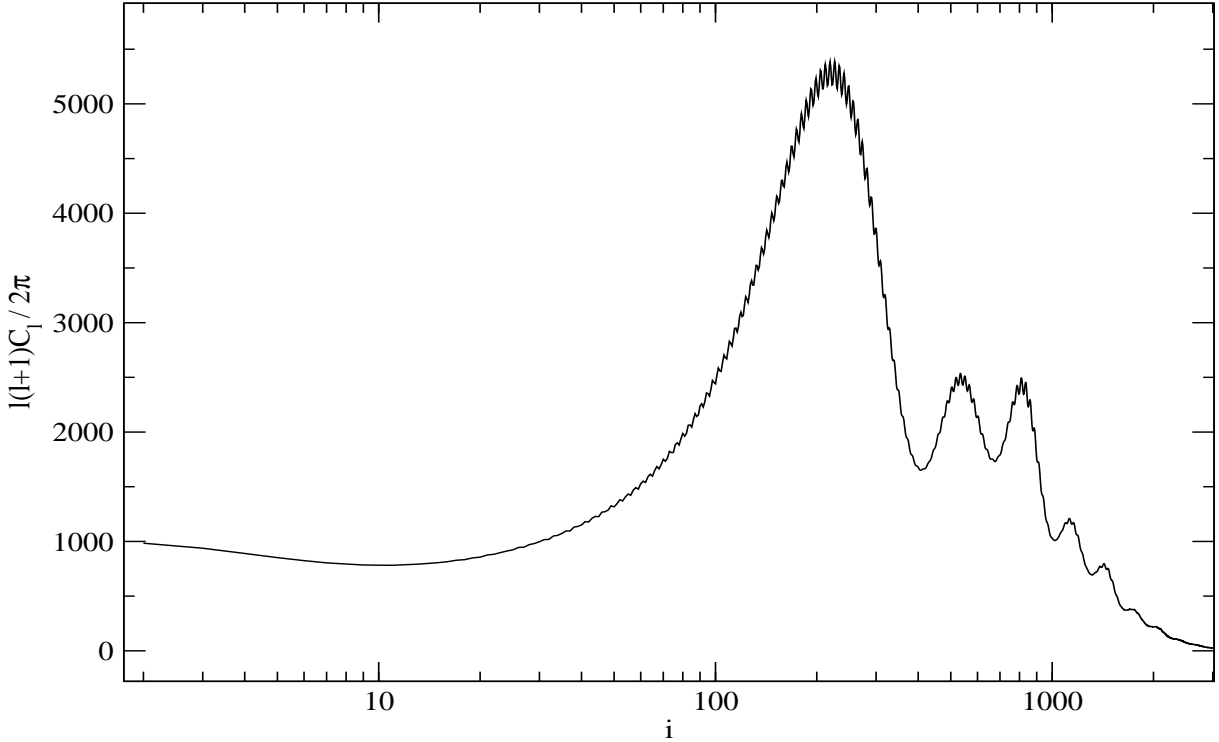}
\caption{Multipole moments in presence of super-imposed trans-Planckian 
oscillations, courtesy of Ref.~\cite{CR}. As discussed in the text, the 
high-frequency super-imposed 
oscillations are damped on large scales and appear at the rise of the first 
Doppler peak.}
\label{fig:tplcl}
\end{figure*}

Before turning to the observational constraints, it is also interesting to 
study how the oscillations in the power spectrum are transferred to the 
multipole moments. In principle, this calculation can only been done 
numerically. There is, however, a regime where one can obtain an approximate 
analytical result. For small $\ell $ (that is to say large angular scales), 
in the limit $\epsilon_1 M_{_{\rm C}}/H\gg 1$ and for $\epsilon_2=-2\epsilon_1$ 
(or $n_{_{\rm S}}=1$), the multipole moments can be expressed as~\cite{Martin:2003sg} 
\begin{eqnarray}
\label{eq:tplcl}
\ell (\ell+1)C_{\ell}&\simeq& \frac{2H^2}{25\epsilon_1\mpl^2}
(1-2\epsilon_1)\Biggl\{1+\sqrt{\pi}\vert x\vert \frac{H}{M_{_{\rm C}}}\ell (\ell+1)
\left(\frac{\epsilon_1M_{_{\rm C}}}{H}\right)^{-5/2}
\nonumber \\ & & \times
\cos \left[\pi \ell+2\frac{M_{_{\rm C}}}{H}
\left(1+\epsilon_1 \ln \frac{\epsilon_1}{a_0Hr_{\rm lss}}\right)+\varphi-\frac{\pi}{4}\right]\Biggr\},
\end{eqnarray}
where $r_{\rm lss}$ is the distance to the last scattering surface. The first conclusion 
that can be drawn from the above equation is that super-imposed oscillations in 
$k$-space are indeed transferred to the $\ell$-space, as the presence of the 
trigonometric function shows. This type of feature is not wiped out by the transfer 
function because the oscillations are present everywhere in $k$-space. Moreover, 
we see that logarithmic oscillations in $k$-space give rise to linear oscillations 
in real space since the argument of the cosine function scales as $\sim \pi \ell$. In 
fact, this last property turns out the be an artefact of the approximations used 
here. Since we are in a situation where the frequency of the oscillations in 
momentum space goes to infinity, one finds oscillations in $\ell$-space 
with frequency $\pi$ which is just the maximum frequency possible given that 
the values of $\ell$ are discrete.

Finally, the amplitude of these oscillations is damped by a factor 
$(\epsilon_1M_{_{\rm C}}/H)^{5/2}$ for very small $\ell$, this effect being compensated 
by $\ell (\ell +1)$ for larger $\ell$. However, since the above formula breaks down in 
the latter regime, it is difficult to reach definite conclusions. One can nevertheless say 
that we expect the super-imposed oscillations to be absent at very large scales and 
to appear only at relatively small scales. These considerations are confirmed in 
Fig.~\ref{fig:tplcl} where multipole moments obtained from a trans-Planckian power 
spectrum have been displayed. In particular, we see that the super-imposed oscillations 
only appear at the rise of the first peak but are damped at very large scales.

Having studied how the trans-Planckian effects affect the CMB anisotropy, let us now 
turn to the question of their possible detection in astrophysical data.

\section{Observational Constraints}

\subsection{The Power Spectrum}

\begin{figure*}
\includegraphics[width=0.9\textwidth,height=.55\textwidth]{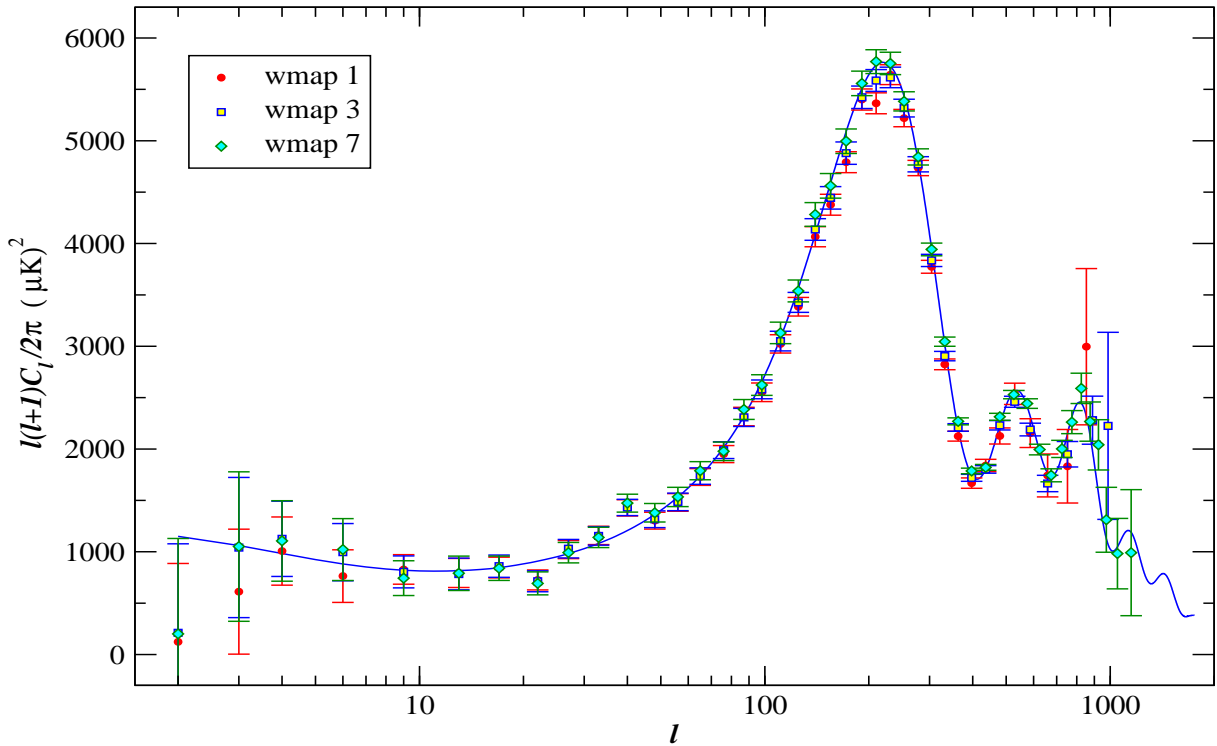}
\caption{Multipole moments measured by the WMAP experiment. The calculation of the 
$C_{\ell}$ depend on the initial power spectrum and, therefore, the corresponding data 
can be used to constrain trans-Planckian physics.}
\label{fig:wmap}
\end{figure*}

We are now ready to discuss the observational constraints on trans-Planckian physics. 
It is conventional to assert that quantum gravity can be probed by studying the propagation 
of high energy cosmic rays. However, this is not the only astrophysical method which 
allows us to constrain trans-Planckian effects. This can also been done with the help 
of the CMB anisotropies that have been precisely measured by the WMAP 
satellite~\cite{Larson:2010gs,Komatsu:2010fb} (see 
Fig.~\ref{fig:wmap}). Here, we briefly review the main techniques that have been used 
to address this question~\cite{Martin:2003sg,Martin:2004iv,Martin:2004yi} (for related 
works in the context of inflation without trans-Planckian effects, see Refs.~\cite{Martin:2006rs,Lorenz:2007ze,Lorenz:2008je,Martin:2010kz,Martin:2010hh}). The exploration of the parameter space corresponding to Eq.~(\ref{eq:powerspectrum}) is usually done using Monte Carlo techniques. Each model 
is computed using a modified version of the {\tt CAMB} code~\cite{Lewis:1999bs} and the 
likelihood is 
estimated using the {\tt COSMOMC} code~\cite{Lewis:2002ah}. 

A first problem met in this kind of analysis is that, in presence of super-imposed 
oscillations, the calculation of one set of multipole moments $C_{\ell}$ becomes 
very time consuming (a few minutes instead of a few seconds). As a consequence, 
it is not possible to explore the full parameter space and the analysis must be 
restricted to the "fast parameter space". This means that the bare cosmological 
parameters are fixed to their best fit values and that the Monte Carlo exploration 
is only performed on the "primordial parameters", namely the overall normalization 
of the spectra, the slow-roll parameters and the parameters describing the oscillations.

The parameters describing the super-imposed oscillations are the amplitude, the frequency 
and the phase. The parameter describing the amplitude is, as discussed above, 
$\vert x\vert H/M_{_{\rm C}}$ and it is convenient to take a uniform prior on this 
parameter in the range $[0,0.45]$. In order to sample the frequency, one can choose 
a uniform prior in $[1,2.6]$ for the parameter $\log(\epsilon_1M_{_{\rm C}}/H)$. 
There is no need to emphasize again the important role played by the fact that 
$\vert x\vert \neq 1$. Taking $\vert x\vert =1$ is not only poorly theoretically 
justified but, from the data analysis point of view, would render the amplitude 
and the frequency completely degenerate, a property that can affect a lot the result 
of the Monte Carlo exploration. Finally, the phase is described by the parameter 
$\psi=2M_{_{\rm C}}(1+\epsilon_1)/H+\varphi$ and one assumes a uniform prior 
in $[0,2\pi]$. 

\begin{figure*}
\includegraphics[width=1.\textwidth,height=.45\textwidth]{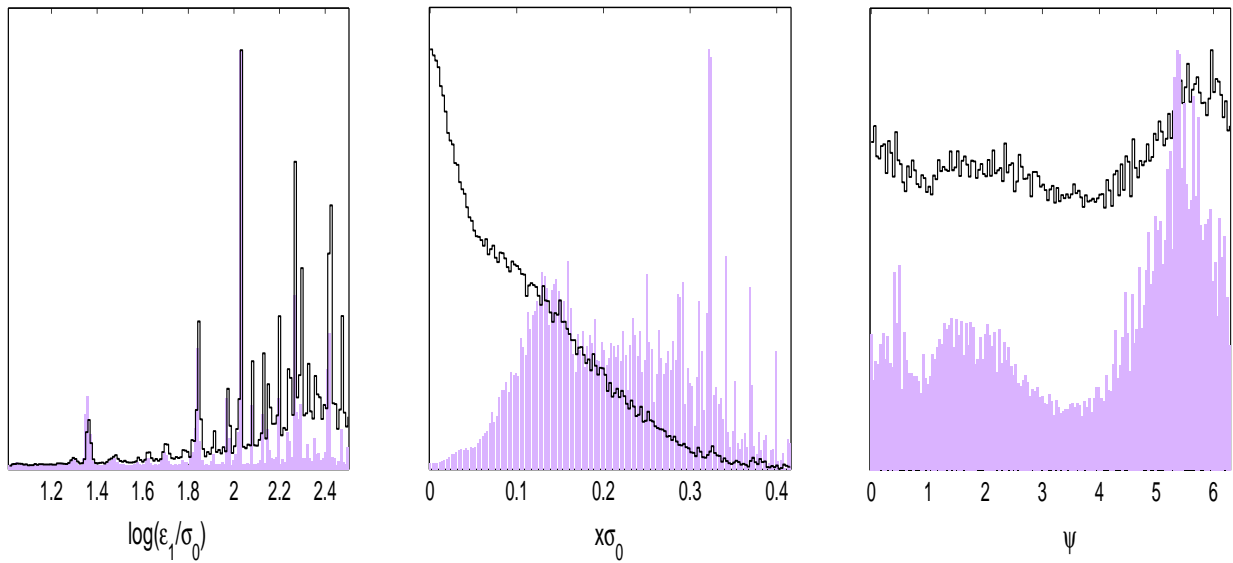}
\caption{Posterior distributions for the three trans-Planckian parameters describing the 
frequency, amplitude and phase of the super-imposed oscillations, courtesy of 
Ref.~\cite{CR}. The parameter 
$\sigma_0$ stands for $H/M_{_{\rm C}}$. The solid black lines represent the marginalized 
distributions while the purple shaded bars represent the mean likelihood.}
\label{fig:post1D}
\end{figure*}

The converged posteriors, marginalised posteriors (solid black line) and mean 
likelihood (purple shaded bars), for the amplitude, frequency and the phase, have 
been plotted in Fig.~\ref{fig:post1D} given the seven years WMAP data~\cite{CR}. The 
marginalised posterior is a quantity which is sensitive not only to the absolute value of the 
likelihood function but also to the volume occupied in parameter space. Therefore, if 
a set of parameters leads to a very good fit but is such that it requires a precise 
fine-tuning of those parameters (that is say, if one slightly detunes the parameters, 
then the likelihood dramatically drops), then its marginalized probability can be small. 
On the other hand, the mean likelihood, as its name indicates, is only sensitive to the 
absolute value of the likelihood function and is independent of volume effects in 
the parameter space which is being considered.

\begin{figure*}
\includegraphics[width=1.\textwidth,height=.45\textwidth]{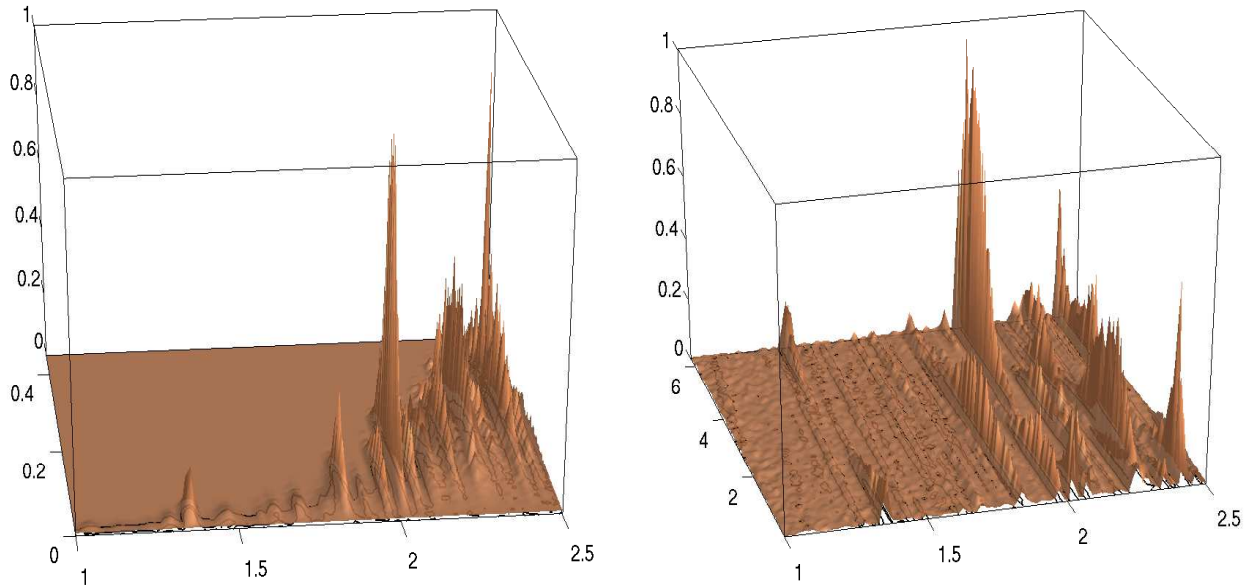}
\caption{Three-dimensional marginalized probability distributions in the amplitude-frequency 
space (left panel) and in the amplitude-phase space (right panel), courtesy of 
Ref.~\cite{CR}.}
\label{fig:post3D}
\end{figure*}

Let us now describe the results displayed in Fig.~\ref{fig:post1D}. The amplitude of the
marginalised probability is peaked over a vanishing value which indicates that the 
vanilla slow-roll power spectrum (with no trans-Planckian oscillations) is still the 
favored model from a Bayesian point of view. One finds that 
$\vert x\vert H/M_{_{\rm C}}<0.26$ at $2\sigma$ confidence level. It is, however, also 
interesting to remark that, as opposed to the marginalised probability, the mean 
likelihood is peaked over a non-vanishing value. This means that the best fit is actually 
a model with trans-Planckian oscillations. One finds that $\Delta \chi ^2\simeq -11$ 
compared to the vanilla slow-roll model. The discrepency between the marginalized 
probability and the mean likelihood is interpreted, in agreement with the above 
discussion, as an indication that volume effects play an important role. In other words, 
if one goes away from the best fit model, the likelihood function quickly decreases which 
means that the best fit only occupies a small volume in parameter space. This analysis 
is confirmed by the frequency posteriors which exhibit a series of peaks at particular 
frequencies. The fact that those peaks are very narrow indicates that it is necessary to 
fine tune the frequency parameter in order to find the best fit, which is totally consistent 
with the fact that the best fit model occupies a small volume and, hence, that the 
marginalised and mean likelihood posteriors differ. Let us add that, in Ref.~\cite{Easther:2004vq}, 
it was shown that, if the data actually contain super-imposed oscillations, then the 
likelihood function should precisely exhibit an oscillatory structure similar to the one 
which seems to emerge from the present analysis (this structure is even more clearly 
visible on the 3D representation of the posteriors plotted in Fig.~\ref{fig:post3D}). This is 
an intriguing fact. Finally, the frequency remains basically unconstrained although 
$\psi \simeq 6$ seems slightly favoured but, in any case, not in a statiscally significant way.

We have emphasized previously that the trans-Planckian effects lead to super-imposed 
oscillations that are logarithmic in $k$. It is therefore interesting to test whether this 
particular scale dependence is favored by the data. This was done in Ref.~\cite{Martin:2006rs} for 
the WMAP3 data. The idea is to postulate the following power spectrum
\begin{equation}
k^3P_\zeta \, = \, P_*\left(\frac{k}{k_*}\right)^{n_{_{\rm S}}-1}
\left\{1-A_\omega\cos\left\{\frac{\omega}{p}\left[
\left(\frac{k}{k_*}\right)^p-1\right]+\psi\right\}\right),
\end{equation}
such that, in the limit $p\rightarrow 0$, one recovers logarithmic oscillations. 
Assuming a uniform prior on $\log p$ in the range $[-5,0.48]$, one obtains at 
$2\sigma$ confidence level that $p<0.68$. In this sense, one can indeed argue 
that the logarithmic structure is special.

The above discussion should however be toned down given the fact that the best fit 
suffers from a severe backreaction problem. Indeed, the best fit is obtained for the 
following values: $\vert x\vert H/M_{_{\rm C}}\simeq 0.13$, 
$\log(\epsilon_1 M_{_{\rm C}}/H)\simeq 2$ and $\log(\epsilon_1)\simeq -3.1$ 
(of course, this last result does not imply a detection of primordial gravitational waves 
since we only explore the fast parameter space, see the discussion before). These 
numbers imply that $H/M_{_{\rm C}}\simeq 10^{-5}$ and $\vert x\vert \sim 10^4$. 
It is interesting to notice that the ratio $H/M_{_{\rm C}}$ has a very natural value and 
that it was indeed very useful to perform the analysis without the assumption 
$\vert x\vert =1$. However, Eq.~(\ref{eq:backreaction}) now reads 
$\vert x\vert H/M_{_{\rm C}}\lesssim 10^{-4.5}$ and the best fit clearly violates this 
inequality by three or four orders of magnitude. Of course, this does not invalidate the 
statistical analysis presented here, it just questions the interpretation of the best fit in 
terms of super-imposed trans-Planckian oscillations.

So, in conclusion, what is the observational status of the trans-Planckian oscillations? 
Clearly, the best model remains the slow-roll vanilla model and there is no detection, 
at a statistically significant level, of super-imposed oscilations. The Bayesian evidence 
for this model has never been computed (it was computed for the first time for different 
slow-roll models only recently in Ref.~\cite{Martin:2010hh}) but it seems pretty clear that its 
estimation would only reinforce this conclusion. Moreover, we do not see any clear 
trend towards a detection as more and more accurate data set are released. Indeed, 
for the first year WMAP data, we had $\vert x\vert H/M_{_{\rm C}} < 0.11$, for the 
three year WMAP data we obtained $\vert x\vert H/M_{_{\rm C}} < 0.38$, which 
is a "better" result, but the result was $\vert x\vert H/M_{_{\rm C}} < 0.26$ for 
the seven year data. There are, however, 
a list of intriguing hints (the best fit is given by a model with super-imposed 
oscillations, presence of narrow peaks in the likelihood function) that seems to indicate 
that, maybe, non trivial features are actually present in the data. Of course, if this is the 
case, it remains to be seen whether there are of primordial origin  and whether they can 
be explained by trans-Planckian physics. But we are of the opinion that these hints are 
a sufficient motivation to carry out the analysis again when the Planck data become available. 

\subsection{Non Gaussianities}

Let us now study how trans-Planckian effects can modify the conventional 
predictions for non-Gaussianities. As is now well-established, the level 
of non-Gaussianity is small in single field inflationary models (with a canonical kinetic term) 
if the initial state is the Bunch-Davies state. Roughly speaking, as shown for the first time in 
Refs.~\cite{Gangui:1993tt,Gangui:1994yr,Wang:1999vf,Gangui:1999vg,Gangui:2000gf} and 
then further elaborated in Refs.~\cite{Maldacena:2002vr}, the $\fnl$ parameter (see the 
definition below) is of the order of the slow-roll parameters \footnote{Although this is very 
often not properly acknowledged 
in the literature, the fact that the non-Gaussianity is small for the class of models mentioned 
above was established in Refs.~\cite{Gangui:1993tt,Gangui:1994yr,Wang:1999vf,Gangui:1999vg} 
much before the publication of Ref.~\cite{Maldacena:2002vr}. The main 
contribution of Ref.~\cite{Maldacena:2002vr} has been to calculate the momentum dependence 
of the bi-spectrum exactly and to introduce efficient techniques which have then permitted the 
calculation of non-Gaussianities for other, more complicated, classes of inflationary scenarios.}. 
This level is too small to be detected even with the Planck satellite (the expected level of 
detection is $\vert \fnl\vert \gtrsim 5$). We have seen, however, that trans-Planckian 
physics, if non-adiabatic, can place the cosmological perturbations in a 
non-vacuum state for which the above-mentioned result no longer holds. It 
is therefore interesting to estimate the level of non Gaussianities if trans-Planckian 
effects are present in the early Universe.

\par

The first calculation in this direction was performed in 
Refs.~\cite{Martin:1999fa,Gangui:2002qc,Gangui:2002qm}. It was done for a non-vacuum 
state with a "preferred direction" in momentum space  such that the three-point correlation 
function does not vanish even in the free theory. In this case, the level of non-Gaussianity 
was found to be extremely small and beyond the level of detectability. We are, however,
more interested in computing the non-Gaussianity for a state characterized by 
Eqs.~(\ref{eq:alpha}) and~(\ref{eq:beta}) which generically describe the trans-Planckian 
modifications. In this case, the free theory is still Gaussian and, therefore, the corresponding 
three-point correlation vanishes. It is only when interactions are taken into account (that is to 
say beyond the Gaussian approximation) that non-Gaussianity can be produced. This 
problem has recently been considered by various authors, see 
Refs.~\cite{Holman:2007na,Meerburg:2009ys,Agullo:2010ws,Ganc:2011dy,Chialva:2011hc}, 
and, in the following, we turn to this question.

\par

A Gaussian distribution has a vanishing three-point correlation function 
and any detection of a non-vanishing signal would therefore rule out 
Gaussian perturbations. For this reason, it is interesting to calculate 
the three point correlation function of the curvature perturbation defined 
by
\begin{eqnarray}
\langle {\cR}(\eta,\vx)\, {\cR}(\eta,\vx)\, 
{\cR}(\eta,\vx)\rangle
&=& \int \f{\d^3 \vka}{(2\,\pi)^{3/2}}\; \int\! \f{\d^3 \vkb}{(2\,\pi)^{3/2}}\;
\int \f{\d^3 \vkc}{(2\,\pi)^{3/2}}\; 
\langle \cR_{\vka} (\eta)\, \cR_{\vkb} (\eta)\, 
\cR_{\vkc} (\eta)\rangle\; {\rm e}^{i\,\l(\vka+\vkb+\vkc\r)\cdot \vx}.
\end{eqnarray}
In the above expression $\cR_{\vk}$ represents the Fourier transform of 
$\zeta $ (and $\vka = {\bm k}_1$ etc.). Let us recall that the curvature perturbation $\zeta $ is related 
to the Mukhanov-Sasaki variable $v$ introduced before 
by $\zeta =v/(\sqrt{2}\Mp a\sqrt{\epsilon_1})$. In the framework of the 
theory of cosmological perturbations of quantum-mechanical origin, it 
is an operator and it can be written as
\begin{equation}
\zeta (\eta, {\bm x})=\int \frac{{\rm d}^3{\bm k}}{(2\pi)^{3/2}}
\left[a_{\bm k}f_{\bm k}(\eta){\rm e}^{i{\bm k}\cdot {\bm x}}
+a_{\bm k}^{\dagger}f_{\bm k}^*(\eta){\rm e}^{-i{\bm k}\cdot {\bm x}}\right].
\end{equation}
Since one has $\zeta_{\bm k}=(a_{\bm k}f_{\bm k}
+a_{\bm k}^{\dagger}f_{\bm k}^*)$, the power spectrum can be expressed as 
[see also Eq.~(\ref{eq:defpsintro})]
\begin{equation}
\label{eq:ngps}
k^3P_{\zeta}\equiv {\cal P}_{\zeta}=\frac{k^3}{2\pi^2}\vert f_{\bm k}\vert^2
=\frac{k^3}{4\pi^2}\frac{\vert v_{\bm k}\vert^2}{\Mp^2a^2\epsilon_1},
\end{equation}
where the quantity $k^3P_{\zeta}$ was also already introduced before, see for instance 
Eq.~(\ref{eq:powerspectrum}). The most general form of $v_{\bm k}$ is given by
\begin{equation}
\label{eq:vdesitter}
v_{\bm k}=\frac{\alpha_{\bm k}}{\sqrt{2k}}\left(1+\frac{1}{ik\eta}\right){\rm e}^{-ik\eta}
+\frac{\beta_{\bm k}}{\sqrt{2k}}\left(1-\frac{1}{ik\eta}\right){\rm e}^{ik\eta}.
\end{equation}
Several comments are in order here. Firstly, the coefficients $\alpha_{\bm k}$ and 
$\beta_{\bm k}$ are of course the same coefficients used to evaluate the trans-Planckian 
power spectrum. The Bunch-Davies vacuum corresponds 
to the choice $\alpha_{\bm k}=1$ and $\beta _{\bm k}=0$. Here, we are interested in a 
situation where $\beta _{\bm k}\neq 0$ and a generic parameterization of the 
deviations from the Bunch-Davies state due to trans-Planckian effects has been 
presented in Eqs.~(\ref{eq:alpha}) and~(\ref{eq:beta}). Secondly, one can check that 
the above expression behaves according to Eq.~(\ref{eq:vsubhubble}) in the sub-Hubble 
regime (in fact, we have slightly changed the conventions - no factor $4\sqrt{\pi}/\mpl$ 
appears anymore - in order to work with notations that are standard in the calculations 
of non-Gaussianities). Thirdly, 
Eq.~(\ref{eq:vdesitter}) corresponds to a de Sitter background. This means that we neglect 
the corrections due to the slow-roll parameters in the mode function (but we took them into 
account when we computed the power spectrum). These corrections would lead to 
sub-leading effects in the calculation of the three point correlation function (for a 
calculation, see for instance 
Refs.~\cite{Chen:2006nt,Chen:2010xka}). Finally, from Eq.~(\ref{eq:vdesitter}), it is 
easy to establish the expression 
of $f_{\bm k}$ that will be used later on. One finds
\begin{equation}
\label{eq:deff}
f_{\bm k}=\frac{i\alpha_{\bm k}H}{2\Mp \sqrt{k^3\epsilon_1}}(1+ik\eta){\rm e}^{-ik\eta}
-\frac{i\beta_{\bm k}H}{2\Mp \sqrt{k^3\epsilon_1}}(1-ik\eta){\rm e}^{ik\eta}.
\end{equation}
We will also need the derivative of the mode function. One obtains
\begin{equation}
\label{eq:deffprime}
f_{\bm k}'=\frac{\alpha_{\bm k}iH}{2\Mp \sqrt{k^3\epsilon_1}}k^2\eta{\rm e}^{-ik\eta}
-\frac{\beta_{\bm k}iH}{2\Mp \sqrt{k^3\epsilon_1}}k^2\eta {\rm e}^{ik\eta},
\end{equation}
where we have neglected terms suppressed by the slow-roll parameters.

\par

As we discussed before, the three point correlation function can only be 
non-vanishing if interactions are taken into account. In this case, the three point 
correlation function in momentum space (or bi-spectrum) can be expressed 
as~\cite{Maldacena:2002vr,Seery:2005wm}
\begin{eqnarray}
\label{eq:inin}
\langle \cR_{\vka}(\eta)\, \cR_{\vkb}(\eta)\, \cR_{\vkc}(\eta)\rangle
= -i\int_{\eta_{\rm ini}}^{\eta_{\rm e}} \d\tau \, a(\tau)
\l\langle\l[\cR_{\vka}(\eta)\, \cR_{\vkb}(\eta)\, 
\cR_{\vkc}(\eta), H_{\rm int}(\tau)\r]\r\rangle,\nonumber \\ 
\end{eqnarray}
with $H_{\rm int}$ being the interaction Hamiltonian while $\eta_{\rm ini}$ is the 
time at which the initial conditions are imposed on the modes when they 
are well inside the Hubble radius. On the other hand, $\eta_{\rm e}$ denotes the time 
when inflation ends. In practice, we always take $\eta _{\rm e}\rightarrow 0$. The 
choice of $\eta_{\rm ini}$ is more tricky in the trans-Planckian context and 
will be discussed in more detail in the following. In order to calculate the 
interaction Hamiltonian, one must compute the action for cosmological perturbations 
at cubic order (the action of the free Gaussian theory is quadratic). A now standard 
calculation gives~\cite{Maldacena:2002vr,Seery:2005wm,Chen:2006nt,Chen:2010xka} 
(a dot means derivative with respect to cosmic time)
\begin{eqnarray}
{\cal S}_{3}[\cR] 
&=& \Mp^2\int \d t\; \d^3{\bm x}\; 
\Biggl[a^3\, \epsilon_{1}^2\, {\cR}\, {\dot {\cR}}^2
+a\,\epsilon_{1}^2\, {\cR}\, (\pa {\cR})^2
-2\,a\, \epsilon_{1}\, {\dot {\cR}}\, 
(\pa^{i}{\cR})\, (\pa_{i}\chi)
\nonumber \\
& &+\,\frac{a^3}{2}\,\epsilon_{1}\, {\dot \epsilon_{2}}\, 
{\cR}^2\, \dot{\cR}
+\frac{\epsilon_{1}}{2a}\, (\partial^{i}\cR)\, 
(\pa_{i}\chi)\, (\pa^2 \chi)
+\f{\epsilon_{1}}{4\,a}\, (\pa^2{\cR})\, (\pa \chi)^2
+ {\cal F}\l(\f{\delta {\cal L}_{2}}{\delta \cR}\r)\Biggr],
\label{eq:cubicaction}
\end{eqnarray}
where $\delta {\cal L}_{2}/\delta\cR$ denotes the variation of the 
second order action with respect to $\cR$, and is 
given by
\begin{equation}
\f{\delta {\cal L}_{2}}{\delta\cR}
={\dot \Lambda}+H\, \Lambda 
-\epsilon_{1}\, \partial^2\cR, 
\end{equation}
and the quantities $\Lambda$ and $\chi$ are defined by
\begin{equation}
\Lambda \equiv \frac{a^2\dot\phi^2}{2\Mp^2H^2}\dot \cR=a^2\epsilon_1\dot \cR,
\quad \chi \equiv \partial ^{-2}\Lambda .
\end{equation}
The term ${\cal F}(\delta {\cal L}_{2}/\delta \cR)$ introduced before
refers to the following complicated expression
\begin{eqnarray}
\label{eq:termeom}
{\cal F}\l(\f{\delta {\cal L}_{2}}{\delta \cR}\r)
&=& \frac{a}{2}\epsilon_2\left(\f{\delta {\cal L}_{2}}{\delta\cR}\right)\cR^2
+\frac{2a}{H}\left(\f{\delta {\cal L}_{2}}{\delta\cR}\right)\dot \cR\cR
\nonumber \\ 
&+&\f{1}{2aH}
\Biggl\{(\partial^{i}\cR)\; (\pa_{i}\chi)\, 
\l(\f{\delta {\cal L}_{2}}{\delta\cR}\r)
+\,\delta^{ij}\, \l[\Lambda\, (\pa_{i}\cR) 
+(\pa^{2}\cR)\, (\pa_{i}\chi)\r]\,
\nonumber \\ 
&\times & \pa_{j}\l[\pa^{-2}\l(\f{\delta {\cal L}_{2}}{\delta\cR}\r)\r]
+
\frac{\delta^{im}\delta^{jn}}{H}(\pa_{i}\cR)\, (\pa_{j}\cR)\;
\pa_{m}\pa_{n}\l[\pa^{-2}\l(\f{\delta {\cal L}_{2}}{\delta\cR}\r)\r]\Biggr\}.
\nonumber \\
\end{eqnarray}
Moreover, the terms which involves $\delta {\cal
  L}_{2}/\delta \cR$ can be removed by a suitable field redefinition
of~$\cR$ of the following
form~\cite{Maldacena:2002vr,Seery:2005wm,Chen:2006nt,Chen:2010xka}:
\begin{equation}
\cR \rightarrow \cR_n+F(\cR_n).\label{eq:frd}
\end{equation}
With such a redefinition, the interaction Hamiltonian corresponding
to the action ${\cal S}_{3}(\cR)$, and to be used in Eq.~(\ref{eq:inin}), can be 
written in terms of the conformal 
time coordinate $\eta $ (a prime means a derivative with respect to conformal time) as
\begin{eqnarray}
H_{\rm int}(\eta) 
&=& -\Mp^2\int \d^{3}{\bm x}\;
\Biggl[a\, \epsilon_{1}^2\, \cR\, \cR'^2  
+ a\, \epsilon_{1}^2\, \cR\, (\pa \cR)^2 
- 2\, \epsilon_{1}\, \cR'\, (\pa^{i}\cR)\, (\pa_{i} \chi)
\nonumber \\
&+& \frac{a}{2}\, \epsilon_{1}\, \epsilon_{2}'\, 
\cR^2\, \cR'
+ \f{\epsilon_{1}}{2\,a}\, (\pa^{i} \cR)\, 
(\pa_{i}\chi)\, \l(\pa^2 \chi\r)
+ \f{\epsilon_{1}}{4\,a}\, \l(\pa^2 \cR\r)\, 
(\pa \chi)^2\Biggr].\label{eq:Hint}
\end{eqnarray}
The three first terms are second order in the slow-roll parameters 
while the three last ones are third order. Despite the appearence of 
the slow-roll parameters, this expression is exact at third order in 
the perturbations.

\par

Having determined the interaction Hamiltonian, one is now in a 
position where non-Gaussianities can be calculated. For convenience, we 
redefine the bi-spectrum according to
\begin{equation}
\langle \cR_{\vka}(\eta _{\rm e})\, \cR_{\vkb}(\eta _{\rm e})\,
\cR_{\vkc}(\eta _{\rm e})\rangle 
=\f{\l(2\,\pi\r)^3}{\l(2\pi\r)^{9/2}}\, \cG(\vka,\vkb,\vkc)
\, \delta^{(3)}\l(\vka+\vkb+\vkc\r),
\end{equation}
where the delta function ensures momentum conservation. Then,
the quantity $\cG(\vka,\vkb,\vkc)$ can be written as
\begin{eqnarray}
\label{eq:defG}
\cG(\vka,\vkb,\vkc)
&=& \Mp^2
\sum_{C=1}^{6}\; 
\Biggl[f_{\ka}(\eta_{\rm e})\,f_{\kb}(\eta_{\rm e})\,f_{\kc}(\eta_{\rm e})
\cG_{_{C}}(\vka,\vkb,\vkc)
\nonumber \\ & &
+f_{\ka}^{\ast}(\eta_{\rm e})\, f_{\kb}^{\ast}(\eta_{\rm e})
\,f_{\kc}^{\ast}(\eta_{\rm e})
\cG_{_{C}}^{\ast}(\vka,\vkb,\vkc)\Biggr],
\end{eqnarray}
The quantities $\cG_{_{C}}(\vka,\vkb,\vkc)$ with ${\small C}=(1,6)$
correspond to the six terms in the interaction
Hamiltonian~(\ref{eq:Hint}), and are given by~\cite{Maldacena:2002vr}
\begin{eqnarray}
\cG_{1}(\vka,\vkb,\vkc)
&=&2i\int_{\eta_{\rm ini}}^{\eta_{\rm e}} \d\tau \,a^2\, 
\epsilon_{1}^2\, \l(f_{\vka}^{\ast}\,f_{\vkb}'^{\ast}\,
f_{\vkc}'^{\ast}+{\rm two~permutations}\r), \nonumber \\
\cG_{2}(\vka,\vkb,\vkc) 
&=&-2i\int_{\eta_{\rm ini}}^{\eta_{\rm e}} \d\tau a^2\, 
\epsilon_{1}^2\, f_{\vka}^{\ast}\,f_{\vkb}^{\ast}\,
f_{\vkc}^{\ast}\,
\l(\vka\cdot \vkb + {\rm two~permutations}\r),\nonumber \\ 
\cG_{3}(\vka,\vkb,\vkc)
&=&-2i\int_{\eta_{\rm ini}}^{\eta_{\rm e}} \d\tau \; a^2\,
\epsilon_{1}^2\, \biggl[f_{\vka}^{\ast}\,f_{\vkb}'^{\ast}\,
f_{\vkc}'^{\ast} \l(\f{\vka\cdot\vkb}{\kb^{2}}\r)
\nonumber \\ & & 
+ {\rm five~permutations}\biggr], \label{eq:ng-dc}\\
\cG_{4}(\vka,\vkb,\vkc)
&=&i\int_{\eta_{\rm ini}}^{\eta_{\rm e}} \d\tau\; a^2\,\epsilon_{1}\,
\epsilon_{2}'\, \l(f_{\vka}^{\ast}\,f_{\vkb}^{\ast}\,
f_{\vkc}'^{\ast}+{\rm two~permutations}\r),\nonumber \\
\cG_{5}(\vka,\vkb,\vkc)
&=&\frac{i}{2}\int_{\eta_{\rm ini}}^{\eta_{\rm e}} \d\tau\; 
a^2\, \epsilon_{1}^{3}\, \biggl[f_{\vka}^{\ast}\,f_{\vkb}'^{\ast}\,
f_{\vkc}'^{\ast} \l(\f{\vka\cdot\vkb}{\kb^{2}}\r)
\nonumber \\ & &
+ {\rm five~permutations}\biggr], \nonumber \\
\cG_{6}(\vka,\vkb,\vkc)
&=&\frac{i}{2}\int_{\eta_{\rm ini}}^{\eta_{\rm e}} \d\tau\; a^2\, 
\epsilon_{1}^{3}\,
\biggl[f_{\vka}^{\ast}\,f_{\vkb}'^{\ast}\, f_{\vkc}'^{\ast}\, 
\l(\f{\ka^{2}}{\kb^{2}\,\kc^{2}}\r)\, \l(\vkb\cdot\vkc\r)
\nonumber \\ & & 
+\, {\rm two~permutations}\biggr]. \nonumber
\end{eqnarray}
Actually, an additional, seventh term arises due to the field
redefinition~(\ref{eq:frd}), and its contribution to 
$\cG(\vka,\vkb,\vkc)$ is found to be
\begin{equation}
\cG_{7}(\vka,\vkb,\vkc)
=\frac{\epsilon_{2}}{2}
\l(\vert f_{\vkb}\vert^{2}\,\vert f_{\vkc}\vert^{2} 
+ {\rm two~permutations}\r).
\end{equation}
The other terms in Eq.~(\ref{eq:termeom}) do not contribute 
because they all contain a derivative (time derivative and/or 
space derivative) and, at the end of inflation, on large 
super Hubble scales, $\zeta$ is constant.

\par

The last thing which remains to be done is to define 
the $\fnl$ parameter which is conventionaly used in the literature 
to quantify the level of non-Gaussianity. It reads
\begin{eqnarray}
\langle \cR_{\vka} \cR_{\vkb} \cR_{\vkc} \rangle
=(2\pi)^{4+3/2}\left(-\frac{3}{10}\fnl
{\cal P}_{_{\rm S}}^2\right)
\frac{\sum _ik_i^3}{\Pi_ik_i^3}\delta (\vka+\vkb+\vkc),
\end{eqnarray}
where we have neglected the slight scale dependence of ${\cal
  P}_{_{\rm S}}$. This definition is introduced to match the 
formula $\cR(\eta, \vx)=\cR^{\rm G}-3\fnl\l(\cR^{\rm G}\r)^2/5$
which is used to extract observational bounds on the parameter
$\fnl$ (the coefficient $3/5$ comes from the fact that, during 
a matter dominated era, the curvature perturbation and the Bardeen 
potential are related by $\zeta=5\Phi/3$).

\par 

>From the above expression, the full bi-spectrum can be calculated. 
For a slow-roll model with canonical kinetic term, the dominant terms 
are $\cG_1$, $\cG_2$, $\cG_3$ and $\cG_7$. This may differ for other 
type of models. For instance, if slow-roll is violated during a few 
e-folds, then the main contribution comes from the term $\cG_4$, see 
Ref.~\cite{Chen:2006xjb,Chen:2008wn,Martin:2011sn,Hazra:2012yn}. 
If the kinetic term is not standard, then a new vertex 
$\propto \dot{\zeta}^3$ appears, see 
Ref.~\cite{Seery:2005wm,Chen:2006nt,Chen:2010xka}. Here, we do not give 
the full bi-spectrum for a slow-roll 
model since this is not the main goal of this review article. It can 
be found in Refs.~\cite{Maldacena:2002vr,Seery:2005wm,Chen:2006nt,Chen:2010xka} (Notice 
that it was recently shown that this bi-spectrum is not modified by re/pre-heating 
effects in single field inflation, see Ref.~\cite{Hazra:2012kq}. Therefore, it actually 
corresponds to what we should see in the CMB). We just quote the so-called consistency 
relation for the squeezed configuration $k_1\simeq k_2\gg k_3$ which reads
\begin{equation}
\label{eq:consistency}
\fnl ^{\rm sq}=\frac{5}{12}(n_{_{\rm S}}-1)\, ,
\end{equation}
and expresses nicely the fact that the non-Gaussianity is indeed of the 
order of the slow-roll parameter (it is also important to recall that 
the validity of this formula is in fact broader, see for instance Ref.~\cite{RenauxPetel:2010ty}; in the case of ultra solw-roll inflation, it is 
violated but the corresponding solution is in fact instable~\cite{Martin:2012pe}).

\par

Having briefly reviewed the standard situation, let us now turn to 
the evaluation of the bispectrum in a situation where $\beta _\vk\neq 0$. In 
principle, one could just restart from Eq.~(\ref{eq:cubicaction}). However, 
it was noticed in Ref.~\cite{Maldacena:2002vr} that the first three terms in this equation 
(that is the say the three dominant terms, second order in the slow-roll 
parameters) can be reduced to 
\begin{equation}
{\cal S}_{3}[\cR] 
=\Mp^2\int \d t\; \d^3{\bm x}\; 4a^3H\epsilon_1^2\zeta'^2
\partial ^{-2}\zeta '\, ,
\end{equation}
up to higher order terms in the slow-roll parameters and terms involving 
$\delta {\cal L}_2/\delta \zeta$. This can be achieved by performing various 
integrations by part. The calculation is then reduced to the calculation of 
a single vertex. This means that the expression~(\ref{eq:defG}) is still 
valid but that the three vertices $\cG_1$, $\cG_2$ and $\cG_3$ are now 
replaced by a single vertex $\cG_{123}$ given by
\begin{equation}
\cG_{123}=8i\left(\sum_{i=1}^3\frac{1}{k_i^2}\right)
\int_{\eta_{\rm ini}}^{\eta_{\rm e}} \d\tau\; a^3H\, 
\epsilon_{1}^{2}\,
f_{\vka}'^{\ast}\,f_{\vkb}'^{\ast}\, f_{\vkc}'^{\ast}.
\end{equation}
In addition, since we have seen that $\cG_4$, $\cG_5$ and $\cG_6$ are 
in fact sub-dominant in the slow-roll approximation, we only have 
to calculate the above vertex. Using the expressions~(\ref{eq:deff}) 
and~(\ref{eq:deffprime}) for the mode function and its derivative, one 
arrives at
\begin{align}
& \langle \cR_{\vka}(\eta _{\rm e})\, \cR_{\vkb}(\eta _{\rm e})\,
\cR_{\vkc}(\eta _{\rm e})\rangle 
=(2\pi)^{-3/2}\delta\left(\vka+\vkb+\vkc\right)\frac{(-i)H^4}{8\Mp^4\epsilon_1}
\frac{k_1^2k_2^2k_3^2}{\prod_{i=1}^3k_i^3}
\left(\sum_{i=1}^3\frac{1}{k_i^2}\right)
\nonumber \\ & \times
\Biggl[\left(\alpha_{\vka}-\beta_{\vka}\right)
\left(\alpha_{\vkb}-\beta_{\vkb}\right)
\left(\alpha_{\vkc}-\beta_{\vkc}\right)
\int_{\eta_{\rm ini}}^{\eta_{\rm e}} \d\tau
\left(\alpha_{\vka}^*{\rm e}^{ik_1\tau}-\beta_{\vka}^*{\rm e}^{-ik_1\tau}\right)
\bigl(\alpha_{\vkb}^*{\rm e}^{ik_2\tau}
\nonumber \\ &
-\beta_{\vkb}^*{\rm e}^{-ik_2\tau}\bigr)
\left(\alpha_{\vkc}^*{\rm e}^{ik_3\tau}-\beta_{\vkc}^*{\rm e}^{-ik_3\tau}\right)
- \mbox{c.c}\Biggr].
\end{align}
We can check that, if $\beta_\vk=0$ in the above equation, together with the 
vertex $\cG_7$, one recovers the consistency relation~(\ref{eq:consistency}).
If $\beta _\vk\neq 0$ however, interesting new effects appears. For instance, 
typically, the bi-spectrum contains the following term
\begin{align}
\langle \cR_{\vka}(\eta _{\rm e})\, \cR_{\vkb}(\eta _{\rm e})\,
\cR_{\vkc}(\eta _{\rm e})\rangle & \supset
\alpha_\vka^*\, \beta_\vkb^*\, \alpha_\vkc^*
\int_{\eta_{\rm ini}}^{\eta_{\rm e}} \d\tau
\, {\rm e}^{i(k_1-k_2+k3)} 
\nonumber \\ & 
\simeq \frac{1}{i(k_1-k_2+k_3)}{\rm e}^{i(k_1-k_2+k3)}
\biggl\vert_{\eta_{\rm ini}}^0
\end{align}
We see that, in the squeezed limit $k_1=k_2\gg k_3$, there is an enhancement of the signal 
by a factor $k_1/k_3\gg 1$~\cite{Ganc:2011dy}. One can therefore hope to constrain 
excited states (and hence 
trans-Planckian physics) with the help of the future non-Gaussianity measurements.

\par

In fact, it is straightforward to calculate the full bi-spectrum for the 
generic parameterization given by Eqs.~(\ref{eq:alpha}) and~(\ref{eq:beta}). One 
obtains
\begin{align}
& \langle \cR_{\vka}(\eta _{\rm e})\, \cR_{\vkb}(\eta _{\rm e})\,
\cR_{\vkc}(\eta _{\rm e})\rangle 
=(2\pi)^{-3/2}\delta\left(\vka+\vkb+\vkc\right)\frac{H^4}{8\Mp^4\epsilon_1}
\frac{k_1^2k_2^2k_3^2}{\prod_{i=1}^3k_i^3}
\left(\sum_{i=1}^3\frac{1}{k_i^2}\right)
\nonumber \\ & \times
\frac{2}{k_{_{\rm T}}}\biggl\{1-\cos\left(k_{_{\rm T}}\eta_{\rm ini}\right)
-3\vert x\vert \frac{H}{M_{_{\rm C}}}\cos \varphi
+3\vert x\vert \frac{H}{M_{_{\rm C}}}\cos \left(k_{_{\rm T}}\eta_{\rm ini}+\varphi\right)
\nonumber \\ & 
-3\vert x\vert \frac{H}{M_{_{\rm C}}}\frac{k_{_{\rm T}}}{k_1+k_2-k_3}
\cos \varphi
+3\vert x\vert \frac{H}{M_{_{\rm C}}}\frac{k_{_{\rm T}}}{k_1+k_2-k_3}
\cos \left[\left(k_1+k_2-k_3\right)\eta_{\rm ini}-\varphi\right]
\nonumber \\ & 
-3\vert x\vert \frac{H}{M_{_{\rm C}}}\frac{k_{_{\rm T}}}{k_1-k_2+k_3}
\cos \varphi
+3\vert x\vert \frac{H}{M_{_{\rm C}}}\frac{k_{_{\rm T}}}{k_1-k_2+k_3}
\cos \left[\left(k_1-k_2+k_3\right)\eta_{\rm ini}-\varphi\right]
\nonumber \\ & 
-3\vert x\vert \frac{H}{M_{_{\rm C}}}\frac{k_{_{\rm T}}}{-k_1+k_2+k_3}
\cos \varphi
+3\vert x\vert \frac{H}{M_{_{\rm C}}}\frac{k_{_{\rm T}}}{-k_1+k_2+k_3}
\cos \left[\left(-k_1+k_2+k_3\right)\eta_{\rm ini}-\varphi\right]
\biggr\},
\end{align}
where we have defined $k_{_{\rm T}}\equiv k_1+k_2+k_3$. This expression 
is valid at leading order in $\beta _\vk$. When $\vert x\vert \rightarrow 0$, 
one recovers the usual result. Let us notice that, in this case 
$\eta_{\rm ini}$ should be rotated 
to the complex plane and sent to infinity. With this usual procedure which 
singles out the vacuum state of the theory, the term $
\cos\left(k_{_{\rm T}}\eta_{\rm ini}\right)$ also vanishes. 
The above expression is quite complicated but we see that the bi-spectrum 
is enhanced when $\tilde{k}_i\equiv \sum _{i=1}^3k_i-2k_i$ 
vanishes~\cite{Meerburg:2009ys}. In order to have a clearer view 
of the result, it is interesting to consider the squeezed limit again. In 
the limit $k_1=k_2=k\gg k_3$, the above equation takes the form
\begin{align}
\langle \cR_{\vka}(\eta _{\rm e})\, \cR_{\vkb}(\eta _{\rm e})\,
\cR_{\vkc}(\eta _{\rm e})\rangle ^{\rm sq}
& =(2\pi)^{-3/2}\delta\left(\vka+\vkb+\vkc\right)\frac{H^4}{8\Mp^4\epsilon_1^2}
\frac{k^3}{\prod_{i=1}^3k_i^3}
\nonumber \\ & \times
\biggl[\epsilon_1+6\, \epsilon_1\vert x\vert \frac{H}{M_{_{\rm C}}}\frac{k}{k_3}
\cos \left(k_3\eta_{\rm ini}-\varphi\right)\biggr]
\end{align}
This simple expression captures the main features of the result. We 
see that the corrections are proportional to $\vert x\vert H/M_{_{\rm C}}$.
We also see the enhancement in the squeezed limit, proportional to 
$k/k_3\gg 1$. This brings us to the question of the initial time $\eta _{\rm ini}$. 
We have seen that, in the minimal approach, the modes are created at different 
times, when their wavelength equals the new fundamental scale $M_{_{\rm C}}^{-1}$, 
see Eq.~(\ref{eq:definitialtime}). In an inflationary background, this implies 
that $\eta _{\rm ini}=\eta_{k}\simeq M_{_{\rm C}}/(kH)$, that is to say 
the initial time becomes scale dependent. However, when one computes the 
various vertices that contribute to the bi-spectrum, one has to integrate 
the product of three mode functions with different wavenumbers $\vka$, $\vkb$ 
and $\vkc$ from $\eta_{\rm ini}$ to zero, see the previous expression for $\cG_{123}$ for instance. 
Then comes the question of which wavenumber should be considered in the expression 
of $\eta _{\rm ini}=\eta_{k}\simeq M_{_{\rm C}}/(kH)$? Let us notice, however, that 
if the scaling $\eta_{\rm ini}\propto M_{_{\rm C}}/(kH)$ is correct (whatever 
the precise definition of $k$ is), then the 
frequency of the oscillatory term in the bi-spectrum becomes 
$\propto M_{_{\rm C}}/H$ in a way which is very similar to the super-imposed 
oscillations found in the power spectrum, see Eq.~(\ref{eq:powerspectrum}).

\par

Finally, let us remark that once the bi-spectrum has been determined in momentum 
space, one still needs to perform the $2D$ projection in order to obtain 
the $\fnl$ seen in the sky. This is a highly non trivial 
procedure~\cite{Meerburg:2009ys,Ganc:2011dy,Ganc:2012ae}. According to these references this 
could lead to an observable $\fnl$ which, therefore, could constrain trans-Planckian 
physics.

\section{Broader Challenges}

\subsection{UV and IR Cutoff Issues}

Let us now return to some more general issues related to trans-Planckian
physics in cosmology. The trans-Planckian problem for cosmological perturbations
is related to basic challenges to the applicability of effective
field theory in an expanding background \footnote{Similar
issues arise in applications to black hole backgrounds and
to non-gravitational external field background.}.

In free quantum field theory, the starting point
is canonical quantization of the fields. In an 
expanding background (see \cite{BD} for an overview
of quantum field theory in curved space-times), the
basic modes which are quantized are the plane wave modes
which have constant wavelength in comoving coordinates
and whose physical wavelength is hence increasing. 
Even in Minkowski space-time,
quantum field theory requires regularization and 
renormalization in order to obtain finite answers for
observables. In the context of gravity, pure
quantum field theory must break down on wavelengths
smaller than the Planck scale since waves with such
large wavelengths would trigger the collapse of the
background space-time into a gas of black holes.
This problem is intimately related to the cosmological
constant problem which arises when considering quantum
field theory in gravity. Hence, in the presence
of gravity, the ultraviolet cutoff is not
just a computational tool, but it as an issue with
definite physical significance. 

The problem in an expanding background is that the
ultraviolet cutoff is expected to correspond to
a fixed scale in physical coordinates, whereas
the basic modes of a quantum field have constant
comoving wavelength. Thus, as already discussed
in \cite{Weiss}, a fixed physical UV cutoff implies
that the Hilbert space of the quantum field must be
time-dependent. As the universe expands, more and
more new modes are required to describe the physics.

The above UV problem arises in any cosmological
space-time. What is special to inflationary cosmology
(as opposed to the alternatives which we have
mentioned in Section 2) is that these new modes
are inflated to a wavelength which at the current
time is observable.

For massless fields, there are also infrared (IR)
divergences. Since gravitational waves are
massless, potential infrared divergences
must be addressed in cosmology. 

In cosmology, there is a natural scale which separates
UV and IR modes - the Hubble scale $H^{-1}(t)$, where
$H(t)$ is the expansion rate of space at time $t$.
On sub-Hubble scales matter fluctuations oscillate
as they do in flat space-time \footnote{Obviously, the
expansion of space leads to a damping of these
oscillations.}. However, on super-Hubble scales the
oscillations are frozen out and plane wave fluctuation
modes become standing waves whose amplitude evolves in
time as determined by the gravitational background.
Furthermore, a local observer (local in space and
time) has a range of view limited by the Hubble
radius. Hence, the Hubble radius determines
the separation between UV and IR modes.
  
The increasing physical wavelength of the basic
modes of a quantum field also leads to an infrared
problem: In the case of an accelerating universe
such as in inflationary cosmology, the phase space
of infrared modes increases. Even in the alternatives
to inflation which were mentioned in Section 2 this
problem arises. In fact, the increase of the
phase space of infrared modes is a necessary condition
to have a causal theory of structure formation
since we need to make sure that scales which are being
observed in cosmology today (and which were in the infrared
sea, i.e. had a wavelength greater than the
Hubble radius for most of the late time universe) start 
out with a wavelength which is sub-Hubble.
In inflationary cosmology, the phase space of infrared modes
increases during the inflationary phase, in string gas cosmology
it grows at the end of the Hagedorn phase, and in
a bouncing cosmology it grows during the contracting
phase.

Infrared divergences can be ``cured'' by imposing an
infrared cutoff. Such an infared cutoff can be well
justfied on physical grounds: it corresponds to setting
to zero the contribution of modes which are super-Hubble
at the initial time, and whose values would depend on
dynamics before that time. In an expanding universe,
this cutoff must be fixed in comoving coordinates, in
contrast to the UV cutoff which is fixed in physical
coordinates (for a discussion of this point see e.g. 
\cite{XueWei1}). 

Since the phase space of IR modes is increasing relative
to the phase space of UV modes, the magnitude 
of the infrared effects will be time-dependent. A possible
consequence of this will be discussed in the following
subsection.

As mentioned above, the growth of the sea of IR modes occurs
in any causal theory of structure formation. What is special
in inflationary cosmology is that the sea becomes populated
by modes which initially were on trans-Planckian scales. Thus,
in inflationary cosmology the trans-Planckian problem for
cosmological fluctuations and the infrared issues become
coupled, which they are not in string gas cosmology or in
a matter bounce scenario.

The effects of infrared modes on the spectrum of cosmological
perturbations in inflationary cosmology has been studied
recently in a large number of works \cite{IRworks}. A first point
to consider is that the effects of IR modes is different in an
inflationary universe (a cosmology with a finite duration de
Sitter phase) compared to what happens in exact de Sitter
space. As recently studied in \cite{XueWei2}, the functional
form of the IR terms is different in the two cases, the extra
terms arising in an inflationary universe being multiplied
by ``slow-roll parameters" of inflationary dynamics.

As first pointed out in \cite{Unruh2} and later discussed in depth in 
\cite{Ghazal1, Abramo}, it is crucial to measure the effects in terms of a physical
clock as opposed to in terms of a background quantity
which itself obtains corrections from the fluctuations. In the
case of purely adiabatic fluctuations, it turns out that the
effects of IR modes are not physically measurable
\cite{Urakawa, Senatore2}. On the other
hand, there are effects of IR modes of entropy fluctuations
which are physically measurable \cite{XueWei2}.

\subsection{IR Effects}

In the previous subsection we have discussed infrared effects on
cosmological fluctuations. What about infrared effects on the
background? From the point of view of cosmological perturbations,
in the same way that a mode with wavenumber $k^{'}$ can combine
with a mode with wavenumber $k - k^{'}$ to cause a second order
correction to the mode with wavenumber $k$, modes with wavenumbers
$k$ and $-k$ lead to a second order correction to the cosmological
background. We call this a ``back-reaction" effect.

In a long-standing research program, Tsamis and Woodard have been
studying the back-reaction of infrared gravitational waves on the
cosmological background \cite{TW}. They find a secular instability
of the de Sitter background, on the basis of which they proposed
a quantum gravitational model of inflation \cite{TW2}.

In \cite{ABM}, the back-reaction effect of scalar metric fluctuations
in the de Sitter phase of an inflationary cosmology was investigated
(see \cite{RHBrev2} for a review of this method).
The starting point was the following ansatz for metric and matter:
\bea \label{ansatz}
g_{\mu \nu}({\bm x}, t) \, &=& \, g_{\mu \nu}^{(0)}(t) + g_{\mu \nu}^{(1)}({\bm x}, t) \nonumber \\
\varphi({\bm x}, t) \, &=& \, \varphi^{(0)}(t) + \varphi^{(1)}({\bm x}, t) 
\eea
where $g_{\mu \nu}^{(0)}(t)$ and $\varphi^{(0)}(t)$ form the cosmological background
and $g_{\mu \nu}^{(1)}({\bm x}, t)$ and $\varphi^{(1)}({\bm x}, t)$ are the linear cosmological 
perturbations (the linearization parameter is the amplitude of the cosmological
perturbations).

This ansatz does not satisfy the Einstein equations beyond linear order.
There are quadratic corrections to both the background and the fluctuations.
Here we are interested in the corrections to the background (see 
\cite{Martineau} for an analysis of the back-reaction on the fluctuations
using the same formalism). We can absorb the quadratic corrections
to the background metric into a new metric $g_{\alpha \beta}^{(0, {\rm br})}(t)$
which equals the original background metric plus quadratic corrections
which take into account that (\ref{ansatz}) does not solve the Einstein
equations to quadratic order.

To obtain the equations of motion for $g_{\alpha \beta}^{(0, {\rm br})}(t)$ we
insert (\ref{ansatz}) into the full Einstein equations and expand to
quadratic order. To extract the equation of motion for the modified
background metric we take the spatial average of the resulting equations.
There are terms quadratic in the linear metric fluctuations which we move
to the right-hand side of the equations of motion to define an
{\it effective energy-momentum tensor} $\tau_{\mu \nu}$ for cosmological perturbations
(see also \cite{Isaacson, Hartle} where this method was first developed,
albeit in a different context). The equation of motion for  $g_{\alpha \beta}^{(0, {\rm br})}$
which includes the quadratic corrections of first order fluctuations
becomes
\be 
G_{\mu \nu} \left [g_{\alpha \beta}^{(0, {\rm br})} \right] \, = \,
8 \pi G \bigl[ T_{\mu \nu}^{(0)} + \tau_{\mu \nu} \bigr]  \, ,
\ee
where $G_{\mu \nu}$ denotes the Einstein tensor, $T_{\mu \nu}$
is the energy-momentum tensor of matter, and $G$ stands for Newton's
gravitational constant.

In terms of the linear metric and matter fluctuations, the effective energy-momentum
tensor takes the form
\be
\tau_{\mu \nu} \, = \, \left \langle 
 T_{\mu \nu}^{(2)} - \frac{1}{8 \pi G} G_{\mu \nu}^{(2)} \right \rangle \, ,
\ee
where the first term contains the terms quadratic in the linear matter fluctuations,
and the second term those quadratic in the linear metric perturbations. There
are also products of metric and matter fluctuations which appear in the first
term (see \cite{ABM} for the full expressions).

The effective energy-momentum tensor $\tau_{\mu \nu}$ obtains contributions
from perturbations of all wavelengths. The contributions of all Fourier modes add
up linearly. We can thus define an UV and an IR part which consist, respectively,
of the contributions of sub-Hubble and super-Hubble modes. In 
the de Sitter limit of inflation, the UV part is time-independent (this follows both
by symmetry and by explicit computation). On the other hand, due to the increasing
phase space of IR modes, the IR part grows in time and hence dominates in
the late time limit.

We now sketch the evaluation of the IR part of $\tau_{\mu \nu}$. To be specific,
we assume that matter takes the form of a scalar field $\varphi$. We work
in longitudinal gauge in which the metric and matter including fluctuations
take the form:
\begin{eqnarray}
{\rm d}s^2 \, &=& \, a^2[(1 + 2 \Phi) {\rm d}\eta^2 
- (1 - 2 \Phi) {\rm d}{\bm x}^2], \quad \varphi \, = \, \varphi_0 + \delta \varphi \, .
\end{eqnarray}
Note that the metric and matter fluctuations $\Phi$ and $\delta \varphi$, respectively,
are related by the Einstein constraint equations
\be
\delta \varphi \, = \, - \frac{2 V}{V^{\prime}} \Phi \, ,
\ee
where $V(\varphi)$ is the potential energy function of $\varphi$.
To obtain the leading terms contributing to the IR part of $\tau_{\mu \nu}$
we can neglect spatial gradient terms and work to leading order in the 
slow-roll approximation for $\varphi$. With these approximations
the IR part of $\tau_{\mu \nu}$ takes the form of a cosmological
constant with effective energy density
\be
\rho^{({\rm br})} \, = \, \tau_0^0 \, \simeq \, 
\left( 2 \frac{V^{''} V^2}{{V^{'}}^2} - 4 V \right) \langle \Phi^2 \rangle
\, .
\ee
For simple inflationary models [e.g. $V(\varphi) = \frac{1}{2} m^2 \varphi^2$]
\be
\rho^{({\rm br})} \, < \, 0 \nonumber
\ee
and hence the back-reaction of IR modes takes the form of a {\it negative}
contribution to the cosmological constant whose magnitude grows in
time since the phase space of IR modes is increasing.

We may conjecture that the back-reaction of IR modes thus leads to a
dynamical relaxation mechanism for a bare positive cosmological 
constant $\Lambda$ \cite{RHBrev2} (for a recent review on the 
cosmological constant problem, see Ref.~\cite{Martin:2012bt}). We begin with
a large bare $\Lambda$ (in the presence of matter). The presence of
$\Lambda$ will lead to a period of inflation.
This, in turn, will cause a sea of IR modes to build up. 
Then, back-reaction will set in. The effect of IR modes will
lead to an effective cosmological constant which determines
the full background metric which takes the form
\be
\Lambda_{\rm eff}(t) \, = \, \Lambda - |\rho^{({\rm br})}(t)| \, ,
\ee
and shows the onset of an instability. 

Naive extrapolation of the back-reaction effect to large times
would lead to $\Lambda_{\rm eff}(t) = 0$ at  some time $t = t_{\rm IR}$.
However, long before this happens the perturbative approach
will break down. Assuming that the instability which is seen to
leading order survives to a non-perturbative treatment, an
interesting scenario emerges \cite{RHBrev2}: The
back-reaction contribution ceases to increase in magnitude
once $\Omega_{\Lambda}(t) < \Omega_{\rm m}(t)$ (where $\Omega_X$
is the contribution of the energy density of the substance $X$ to 
the critical density). since at that
time the acceleration of space ceases and the infrared sea
ceases to grow. The universe, however, is still expanding,
and hence the matter energy density decreases. Thus,
the effective cosmological constant once again raises its
head. The upshot of this dynamics is that there is a
dynamical fixed point with
\be
\Omega_{\Lambda}(t) \, \sim \, \frac{1}{2} \, .
\ee
More precisely, oscillations of $\Omega_{\Lambda}(t)$ about that 
value are expected. Thus, the back-reaction of long wavelength
cosmological perturbations has the potential of providing
a dynamical relaxation mechanism for a large bare cosmological
constant which explains dark energy and has no coincidence
problem.

Note that the above mechanism is not in conflict with causality
since only modes are considered which are inside the horizon 
\footnote{Recall that in inflationary cosmology the horizon is
larger than the Hubble radius by a factor which exponentially
increases during inflation.}. The effect can be locally described
in terms of a time-dependent change in the spatial curvature
constant $k$ and $\Lambda$ \cite{Lam}.

The key concern is that the above calculations have been performed
as a function of a background time $t$. As conjectured in
\cite{Unruh2} and verified in \cite{Ghazal1, Abramo}, in the case
of purely adiabatic fluctuations (the energy densities of
all components of matter being proportional) the leading IR
back-reaction effect discussed above can be entirely absorbed
by a second order time-reparametrization. In other words, if
we calculate observables such as the local Hubble expansion
rate as a function of a physical clock variable as opposed to
the background time $t$, there is no leading order effect of
the IR fluctuations. For adiabatic fluctuations, the only
physical clock is the matter field $\varphi$, and we find
\cite{Ghazal1}
\be
H^2(\varphi, \Phi) \, = \, H^2(\varphi, 0) \, .
\ee
However, in the presence of multi-field systems
with entropy fluctuations, it can be shown that
the leading IR back-reaction effects are physically
measurable \cite{Ghazal2} (see also \cite{Urakawa2}). 
If $\chi$ is an entropy
field (e.g. a field which represents the temperature
of the CMB), then
\be
H^2(\chi, \Phi) \, \neq \, H^2(\chi, 0) \, .
\ee
Thus, there appears to be evidence for an instability of the de Sitter
phase of inflationary cosmology. What about de Sitter space itself?
We will end with a very brief literature survey of work on this topic.

\subsection{Stability of de Sitter}

De Sitter space-time is a classical solution of Einstein's field
equations in the presence of a positive cosmological constant.
According to the classical ''no-hair" theorems \cite{Starob3, Wald}
the expanding branch is classically stable (the contracting
branch is classically unstable because of the growth of
fluctuations in such a phase). 

As discussed in the previous subsection, there is perturbative
evidence that in the presence of matter and entropy fluctuations
even the expanding branch of de Sitter space is semi-classically 
unstable. This question
has been studied for a long time. Based on studies of the
renormalization of the energy-momentum tensor of a massless
scalar field in de Sitter space there have been claims of
instability \cite{Ford, Traschen}. There are also claims of
instability due to particle production \cite{Myrvold, Emil1}, due
to a thermodynamic instability \cite{Emil2} and due to a
conformal anomaly \cite{Emil3}.  The work of \cite{TW}
mentioned in the previous subsection supports the claim
of semi-classical instability.

What about non-perturbative statements? In the context of
four space-time dimensional gravity there have been
long-standing claims that de Sitter space is unstable
\cite{Polyakov} (see also \cite{others}). However, these
claims have been disputed in \cite{Marolf}.

There has been some interesting recent work on the
stability issues of de Sitter space. In the context of
three space-time dimensional pure gravity it has
been shown that the method used to construct the
partition function of quantum gravity which works in
anti-de-Sitter space breaks down in the presence
of a positive cosmological constant \cite{Castro}.
In the context of the higher spin theory limit of string
theory is has also been shown that de Sitter space
does not seem to be a solution at the quantum level
\cite{Dio}.
 
\section{Conclusions}

We have seen that the exponential expansion of space
during the expanding branch of de Sitter space-time,
and consequently also during an inflationary phase of
an expanding FRWL universe, leads to important
trans-Planckian issues. In particular, fluctuation
modes which are being probed in current cosmological
observations originate with wavelengths smaller than the
Planck length at the onset of the phase of exponential
expansion. Since the physics on trans-Planckian
scales is unkown, this raises conceptual questions
regarding to the robustness of the usual calculations
in inflationary cosmology.

One aspect of this problem, namely the fact that in an
expanding space comoving modes continuously
cross the boundary of the trans-Planckian zone of
ignorance, is common to all expanding cosmologies.
What is special to inflationary cosmology is that
scales which are currently explored in cosmological
observations emerge from this trans-Planckian sea.
This is not the case in the two other cosmological
paradigms mentioned earlier, namely the matter bounce
scenario and string gas cosmology.

These issues have been discussed in Section 2 of this
article. Consequences for cosmological observations have
been discussed in Section 3 and 4. Possibly the most
important point is that due to the accelerated increase in
the wavelength of scales, trans-Planckian physics
can in fact be tested with current cosmological observations.
We have seen that
under the assumption that at some initial time all modes
are in their local vacuum state, then oscillations in the
spectrum of CMB anisotropies are induced. The strongest
constraint on the amplitude of these oscillations comes
from back-reaction considerations, namely from demanding
that the produced particles do not destabilize the 
inflationary background. Such oscillations can be
searched for in current CMB anisotropy data. With current
results, a vanilla inflationary model without oscillations is
still an excellent fit to the data.

The accelerated expansion of space during an inflationary
period leads to more conceptual problems. The increase
of the wavelength of fixed comoving modes relative to the
Hubble radius leads to a rapid increase of the phase space
of such infrared (IR) modes. This may - in the presence of
entropy fluctuations - lead to large IR effects, possibly even to
a de-stabilization of the de Sitter phase.

\section*{Acknowledgements}

We would like to thank C.~Ringeval for having shared his 
numerical results with us. The work of R.~B. is supported in part 
by an NSERC Discovery Grant and by funds from the Canada Research 
Chairs Program. J.~M. would like to thank L.~Sriramkumar for useful 
discussions and the University of Tokyo (Research Center for the Early 
Universe) where part of this work has been done.

\end{document}